\documentclass[sigconf, 9pt]{acmart}

\AtBeginDocument{%
  }

\usepackage{balance}
\usepackage{subcaption}
\usepackage{array}
\usepackage{multirow}
\usepackage{amsmath}
\usepackage{comment}
\usepackage{url}
\usepackage{todonotes} 
\usepackage{fontawesome5}
\usepackage{mathtools}

\usepackage{pgfplots}
\pgfplotsset{width=\columnwidth,compat=1.9}
\usetikzlibrary{patterns}

\usepackage{tikz}

\usepackage{tabularx}
\usepackage{hyperref} 

\newcommand{\bftsmart}{\textsc{BFT-SMaRt}}
\newcommand{\ourProtocol}{\textsc{Mercury}}
\newcommand{\ourProtocolShort}{\textsc{Mercury}}
\newcommand{\aware}{AWARE}
\newcommand{\wheat}{WHEAT}

\usepackage{algorithmic}
\usepackage[linesnumbered,vlined,boxed,commentsnumbered, ruled]{algorithm2e} 
\usepackage{amsmath,amsthm,amsfonts}
\usepackage[most]{tcolorbox}
\usepackage{enumitem}

\newtheorem{theorem}{Theorem}
\newtheorem{lemma}{Lemma}
\newtheorem{proposition}{Proposition}

\definecolor{dartmouthgreen}{rgb}{0.05, 0.5, 0.06}
\definecolor{duke-blue}{rgb}{0.0, 0.0, 0.61}

\definecolor{dark-blue}{rgb}{0.18, 0.33, 0.59}
\definecolor{light-blue}{rgb}{0.56, 0.67, 0.86}

\definecolor{dark-red}{rgb}{0.49, 0, 0}
\definecolor{light-red}{rgb}{0.75, 0, 0}

\definecolor{dark-orange}{rgb}{0.77, 0.35, 0.07}
\definecolor{light-orange}{rgb}{0.92, 0.49, 0.19}

\definecolor{dark-yellow}{rgb}{1, 0.75, 0}
\definecolor{light-yellow}{rgb}{1, 0.85, 0.4}

\definecolor{dark-green}{rgb}{0.66, 0.81, 0.56}
\definecolor{light-green}{rgb}{0.77, 0.88, 0.71}

\definecolor{db1}{rgb}{0.08,0.14,0.25}
\definecolor{lb1}{rgb}{0.12,0.22,0.39}

\definecolor{db2}{rgb}{0.23,0.3,0.44}
\definecolor{lb2}{rgb}{0.21,0.32,0.56}

\definecolor{db3}{rgb}{0.29,0.48,0.64}
\definecolor{lb3}{rgb}{0.51,0.58,0.73}

\definecolor{db4}{rgb}{0.55,0.67,0.86}
\definecolor{lb4}{rgb}{0.71,0.78,0.91}

\definecolor{links-blue}{HTML}{0071BB}

\newcommand*\serverRule[1]{\tikz[baseline=(char.base)]{
		\node[shape=circle,draw,inner sep=0.25pt, fill=dark-blue, color=dark-blue, text=white] (char) {\textbf{#1}};}}

  \newcommand*\clientRule[1]{\tikz[baseline=(char.base)]{
		\node[shape=circle,draw,inner sep=0.25pt, fill=light-blue, color=light-blue, text=white] (char) {\textbf{#1}};}}

  \newcommand*\auditRule[1]{\tikz[baseline=(char.base)]{
		\node[shape=circle,draw,inner sep=0.25pt, fill=duke-blue, color=duke-blue, text=white] (char) {\textbf{#1}};}}

\hypersetup{
  colorlinks   = true,    
  urlcolor     = links-blue,    
  linkcolor    = links-blue,    
  citecolor    = links-blue      
}

\begin{document}

\title{\emph{Chasing the Speed of Light}: \\ Low-Latency Planetary-Scale Adaptive Byzantine Consensus}



%

\author{Christian Berger}
\affiliation{%
  \institution{University of Passau}
  \city{Passau}
  \country{Germany}}
\email{cb@sec.uni-passau.de}

\author{Lívio Rodrigues}
\affiliation{%
  \institution{LASIGE, Faculdade de Ciências, Universidade de Lisboa}
  \city{Lisboa}
  \country{Portugal}}
\email{lgrodrigues@ciencias.ulisboa.pt}

\author{Hans P. Reiser}
\affiliation{%
  \institution{Reykjavik University}
  \city{Reykjavik}
  \country{Iceland}}
\email{hansr@ru.is}

\author{Vinícius Cogo}
\affiliation{%
  \institution{LASIGE, Faculdade de Ciências, Universidade de Lisboa}
  \city{Lisboa}
  \country{Portugal}}
\email{vvcogo@ciencias.ulisboa.pt}

\author{Alysson Bessani}
\affiliation{%
  \institution{LASIGE, Faculdade de Ciências, Universidade de Lisboa}
  \city{Lisboa}
  \country{Portugal}}
\email{anbessani@ciencias.ulisboa.pt}

\renewcommand{\shortauthors}{C. Berger, L. Rodrigues, H.P. Reiser, V. Cogo and A. Bessani}
\begin{abstract}
Blockchain technology sparked renewed interest in planetary-scale Byzantine fault-tolerant (BFT) state machine replication (SMR). 
While recent works predominantly focused on improving the scalability and throughput of these protocols, few of them addressed latency.
We present \ourProtocol{}, a novel transformation to autonomously optimize the latency of quorum-based BFT consensus. \ourProtocol{} employs a dual resilience threshold that enables faster transaction ordering when the system contains few faulty replicas.
\ourProtocol{} allows forming \textit{compact quorums} that substantially accelerate consensus using a smaller resilience threshold. Nevertheless, \ourProtocol{} upholds standard SMR safety and liveness guarantees with optimal resilience, thanks to its judicious use of a dual operation mode and BFT forensics techniques. Our experiments spread tens of replicas across continents and reveal that \ourProtocol{} can order transactions with finality in less than $0.4$s, half the time of a PBFT-like protocol (optimal in terms of number of communication steps and resilience) in the same network.
Furthermore, \ourProtocol{} matches the latency of running its base protocol on theoretically optimal internet links (transmitting at $67\%$ of the speed of light). 
\end{abstract}


\settopmatter{printacmref=false}
\setcopyright{none}
\renewcommand\footnotetextcopyrightpermission[1]{}
\pagestyle{plain}

\maketitle

\section{Introduction}

State machine replication (SMR) is an approach to tolerate faults in distributed systems by coordinating client interactions with a set of $n$ independent replicas~\cite{schneider1990implementing}.
Recently, many scalable (BFT) SMR protocols have emerged for blockchain infrastructures, such as HotStuff~\cite{hotstuff19}, SBFT~\cite{gueta2019sbft}, Tendermint~\cite{cason2021design}, Mir-BFT~\cite{stathakopoulou2019mir}, RedBellyBC~\cite{crain2021red}, Kauri~\cite{neiheiser2021kauri}, IA-CCF~\cite{shamis22iaccf}, and the Dumbo family~\cite{boltdumbo,dumbong}. 
These protocols employ some dynamically elected leader~\cite{hotstuff19, gueta2019sbft, cason2021design, neiheiser2021kauri,sui2022marlin}, use multiple leaders~\cite{stathakopoulou2019mir, alqahtani2021bigbft}, or are leaderless~\cite{crain2021red, antoniadis2021leaderless, dumbong,pace}.


\begin{figure}[!t]
    \centering
  \begin{subfigure}[h]{0.8\columnwidth}
  \centering
    \includegraphics[width=1\columnwidth]{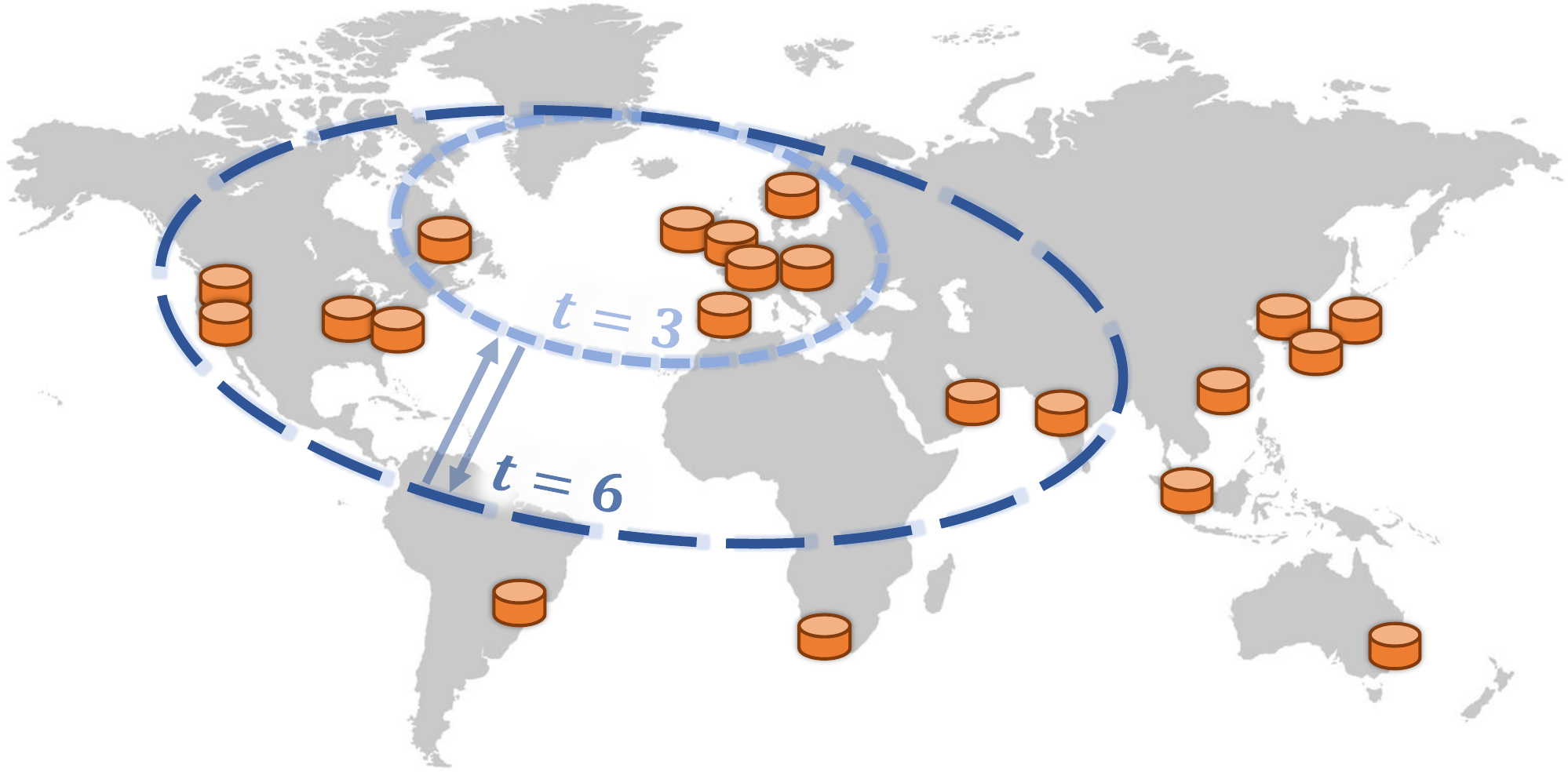}
    \caption{Weighted quorums sizes with {\color{dark-blue} $t=6$} and {\color{light-blue} $t=3$}.}
    \label{fig:aws21-WHEAT-quorums}
  \vspace{2mm}
  \end{subfigure}
  \begin{subfigure}[h]{0.39\columnwidth}
         \begin{tikzpicture} 
    \begin{axis}[ 
     font= \small,
     ylabel={Latency [ms]}, 
     xlabel={Threshold $t$}, 
     xticklabels from table={data/consensus-latencies.txt}{t},    
     ybar=2pt,  
     bar width=6pt,
    height=3.2cm,
       xtick=data, 
       ytick = {0, 50,100,150,200,250, 300},
        ymin=0,
        ymax=300,
        xmin=2.5,
        xmax=6.5,
    ymajorgrids=true,
    yminorgrids=true,
    minor grid style={dashed,gray!10},
    minor tick num=1,
    legend style={at={(1,1.3)},
    legend cell align=left
    }
    ] 
\addplot 
[draw = black!80!white, 
fill = black!60!white]   
table[ 
  x=t, 
  y=latency    
  ] 
{data/consensus-latencies.txt}; 
\end{axis} 
\end{tikzpicture} 
      \caption{Consensus latency \\\emph{vs.} resilience threshold.}
    \label{fig:aws21-consensus-latency}
  \end{subfigure}
  \begin{subfigure}[htb]{0.6\columnwidth}
  \begin{tikzpicture} 
    \begin{axis}[ 
    font= \small,
     ylabel={Latency [ms]}, 
     xticklabels from table={data/client-latencies-intro.txt}{region},   
        x tick label style={rotate=30,anchor=east,  xshift=10pt, yshift=-4pt,     font= \small},
     ybar=2pt,  
     bar width=6pt,
    height=3.2cm,
       xtick=data, 
       ytick = {0, 100,200,300,400,500,600},
        ymin=0,
        ymax=600,
        xmin=0.5,
        xmax=4.5,
    ymajorgrids=true,
    yminorgrids=true,
    minor grid style={dashed,gray!10},
    minor tick num=1,
    legend style={at={(1, 1.2)},
    legend columns = 5,
    legend cell align=left
    }
    ] 
      \addplot 
       [draw = dark-blue,
        fill = dark-blue!80!white]   
        table[ 
          x=regionNr, 
          y=t6   
          ] 
      {data/client-latencies-intro.txt}; 
     \addlegendentry{ $t=6$  \  }; 
   
            \addplot 
       [draw = light-blue, 
        fill = light-blue!80!white,
        postaction={pattern=north east lines,pattern color=light-blue!80!white}]   
        table[ 
          x=regionNr, 
          y=t3 
          ] 
      {data/client-latencies-intro.txt}; 
       \addlegendentry{$t=3$ \ };     
    \end{axis} 
\end{tikzpicture} 
\caption{End-to-end transaction latencies\\observed by clients in different regions.}
\label{fig:aws21-request-latency}
  \end{subfigure}
    \caption{Weighted quorums composition and resulting BFT SMR latency for different resilience thresholds ($t$) in our $n=21$ setup (see details in \S\ref{sec:evaluation}).}
    \label{fig:intro-latencies}
    \vspace{-3mm}
\end{figure}

Nevertheless, the consensus in all these cases requires communication involving a quorum of replicas under the assumption that the adversary controls no more than a \emph{fixed resilience threshold} of $t=\lfloor \frac{n-1}{3} \rfloor$ replicas.
Often, the quorum size for proceeding to the next protocol stage depends on this threshold, a Byzantine $t$-\emph{dissemination quorum} with $\lceil \frac{n+t+1}{2} \rceil$ replicas~\cite{malkhi1998byzantine}.
This size equals roughly $\frac{2}{3}$ of all replicas if an \emph{optimal} resilience threshold is used.
Two challenges arise in optimizing end-to-end client latency for geo-replicated or planetary-scale systems like permissioned blockchains (e.g., \cite{androulaki2018hyperledger,crain2021red}) with tens of nodes distributed worldwide.
First, theoretical lower bounds define that at least three communication steps are required for reaching consensus without giving up the optimal resilience~\cite{martin2006fast,kuznetsov2021revisiting}.
Second, there are physical limits that bound link transmission speed to a fraction of the speed of light (e.g., $0.67c$~\cite{kohls2022verloc}).
Contrarily, improving throughput is a much more popular objective that can be achieved by parallelizing/distributing tasks (e.g.,~\cite{crain2021red,pace}), improving bandwidth usage (e.g.,~\cite{scalabilitymadesimple,probft}), or simply by using a better infrastructure (e.g., better network links).
Nonetheless, globally ordering transactions in a fraction of a second is still far from reality for existing systems~\cite{gramoli23diablo}, making near-instantaneous confirmation of transactions a missing usability feature of blockchains.

\subsection{Smaller Quorums for Better Latency}

We advocate that using smaller quorums of closer replicas can significantly decrease SMR latency~\cite{junqueira2007classic,wheat}.
The challenge lies in ensuring these faster, smaller quorums intersect in sufficiently many replicas with all other quorums of the system.
Such quorums can be built using \emph{weighted replication}, giving faster replicas more voting power.

Figure~\ref{fig:intro-latencies} illustrates how a geo-replicated system can progress faster with smaller quorums.
It considers a weighted system~\cite{wheat} with $n=21$ replicas dispersed across all $21$~AWS regions (see Figure~\ref{fig:aws21-WHEAT-quorums}).
When configured for maximum resilience, this system tolerates up to $t=6$ Byzantine replicas 
with $\Delta = 2$ spare replicas. 
The smallest weighted consensus quorum~$Q_6$ contains $13$ replicas (see \S\ref{sec:wheat} for details on the calculations), only one replica less than using non-weighted replication.
Alternatively, when configured for $t=3$ failures, the smallest weighted quorum~$Q_3$ contains only $7$ replicas, with $\Delta = 11$.
This quorum can comprise the nearest replicas that can exchange votes with each other faster, accelerating the consensus protocol stages (see Figure~\ref{fig:aws21-consensus-latency})
and resulting in end-to-end latency improvements around the globe (see Figure~\ref{fig:aws21-request-latency}).

\subsection{Challenges and the Big Picture}

The problem in using a lower resilience bound $t_\mathit{fast} < t$ is that an adversary controlling $f$ replicas (with \mbox{$t_\mathit{fast} < f \leq t$}) can equivocate.
It means the adversary can convince \textit{two correct} replicas to decide \textit{different batches} of transactions for the same consensus instance, as quorums for the lower threshold $t_\mathit{fast}$ do not necessarily overlap in at least one correct replica.

A key insight of our work is the innovative use of \emph{BFT protocol forensics}~\cite{shamis22iaccf,bftforensics} as a defense against Byzantine attackers rather than post-incident forensics.
Traditional BFT protocol forensics relies upon clients to detect conflicting values based on replicas' logged messages and pinpoint the equivocating replicas.
In our solution, we impose the responsibility of detecting faulty replicas on all correct replicas, enabling the autonomous detection and removal of equivocating parties.
In a system tolerating up~to~$t_\mathit{fast}$ faulty replicas, audits can detect agreement violations and identify $t_\mathit{fast}+1$ faulty replicas if there are no more than $2t_\mathit{fast}$ faulty replicas~\cite{bftforensics}. 
Using $t_\mathit{fast} = \lceil \frac{t}{2} \rceil$ guarantees that audits are always supported for up to $t$ faulty replicas.
In our approach, the system recovers from violations by expelling the detected equivocators and rolling back the divergent decisions of correct replicas to a consistent state.
Continuous auditing is important not only as a recovery mechanism but also as a \emph{deterrent to attacks} since any perpetrator is identified and expelled from the system.

The ability to roll back decisions on replicas may lead to transaction outcomes observed by clients being undone, affecting transaction finality and durability.
Consequently, we must modify the matching replies requirements on clients to ensure they can know when an operation is \emph{finalized} in the system, ensuring consistency (i.e., linearizability~\cite{herlihy1990linearizability}) and liveness as in standard SMR~\cite{pbft}.
Another challenge we address is deriving the exact number of matching replies a client needs to expect to finalize a submitted transaction when consensus agreement violations are possible.

Minimizing consensus latency but letting clients wait for \emph{more-than-usual} replies from all over the world to preserve linearizability counteracts our goal of reducing the end-to-end request latency.
For this reason, we extend the BFT SMR programming model with \emph{Byzantine correctables}, empowering clients with incremental consistency guarantees~\cite{guerraoui2016incremental} and enabling the early confirmation of submitted transactions.

Lastly, we need to consider \emph{SMR liveness}, as requests issued by correct clients need to be eventually completed.
This property can be endangered if the protocol operates with an optimistic threshold $t_\mathit{fast}$ and there are $f>t_\mathit{fast}$ Byzantine replicas that stay silent, i.e., do not reply to the client or participate in consensus quorums.
To deal with this scenario, we employ the idea of having two modes of operation:
If the system blocks or equivocates, we stop the execution of \ourProtocol's fast mode and resume the execution of the standard protocol tolerating $t$ Byzantine replicas.

Our experimental evaluation with up to $51$ replicas around the globe shows that \ourProtocol{} can order transactions with finality in less than $0.4$s, which is half of the time required for \bftsmart{}~\cite{bftsmart} (which implements a PBFT-like protocol) in the same network.
Interestingly, our observed latencies are close to the theoretical optimum for \bftsmart{}, considering the physical location of replicas and links transmitting at $\frac{2}{3}$ of the speed of light, which is accepted as the upper bound on data transmission speed for the internet~\cite{cangialosi2015ting,kohls2022verloc}.
Further, we achieve consensus latencies $4\times$ smaller than recent results reported in state-of-the-art protocols targeting low-latency in similar environments (e.g.,~\cite{babel2024mysticeti,boltdumbo}).

\subsection{Contributions}
\ourProtocol{} shows how to obtain a threshold-adaptive BFT protocol that strives for continuous self-optimi\-zation during runtime by tuning the resilience threshold utilized in consensus quorums.
This protocol significantly reduces the latency in planetary-scale BFT SMR in the expected common case with few failures.
In summary, we claim the following contributions:

\begin{itemize}
    \item We study how to detect malicious behavior under an underestimated threshold $t_\emph{fast}$ by periodically auditing the system, removing faulty replicas, and repairing the correct replicas' state after an agreement violation.
    \item We show that it is possible to preserve the usual SMR guarantees, \emph{linearizability} and \emph{termination}, under the larger resilience threshold $t$, even if the agreement quorums are formed using a smaller threshold $t_\emph{fast}<t$.
    \item We introduce \emph{Byzantine correctables} to allow for client-side speculation, thus enabling a client application to minimize the observed transaction latency even further by selecting the desired consistency level of their transactions.
    \item We present an extensive evaluation of \ourProtocol{} in real and simulated networks, characterizing the end-to-end latency improvements of the proposed approach.
    \item We show that the principles underlying \ourProtocol{} can generalize to other quorum-based BFT protocols such as HotStuff~\cite{hotstuff19}, resulting in a greater relative latency reduction for HotStuff.
\end{itemize}

We prove the correctness of \ourProtocol{} in the appendix of this paper, and all the code employed in our experiments is available online~\cite{mercurycodebase}.



 \begin{figure*}[t]
\centering
\begin{subfigure}[t]{.3\textwidth}
\centering
   \includegraphics[width=1\columnwidth]{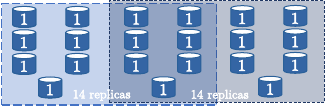}
   \caption{\emph{Egalitarian ($t=6$)}: All quorums have the same size of  $\lceil\frac{n+t+1}{2}\rceil$ replicas.}
   \label{fig:WHEAT_egalitarian}
\end{subfigure}
  \hfill
\begin{subfigure}[t]{.32\textwidth}
\centering
      \includegraphics[width=1\columnwidth]{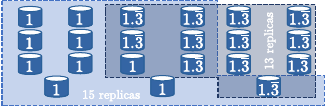}
   \caption{\emph{Weighted ($t=6$)}: Every quorum contains at least $2t+1$ and at most $n-t$ replicas.}
\label{fig:WHEAT_weighted} 
\end{subfigure}
\hfill
\begin{subfigure}[t]{.31\textwidth}
\centering
      \includegraphics[width=1\columnwidth]{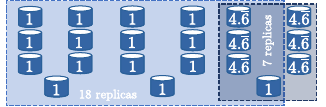}
   \caption{\emph{Weighted ($t=3$)}: Quorums can become \emph{small} by intersecting in \emph{fast} replicas.}
\label{fig:WHEAT_weighted_fast} 
\end{subfigure}
\caption[]{Overview over BFT quorum systems for $n=21$ replicas.
}
\label{fig:WHEAT_quorums} 
\end{figure*}

\section{Background}
\label{sec:background}

In this section, we first highlight some fundamental concepts of BFT SMR and then revisit  \textit{weighted quorums} in BFT replication.

\subsection{Byzantine State Machine Replication}
\label{sec:state_machine_replication}

Assuming a deterministic replicated service in which all replicas start at the same state~\cite{schneider1990implementing}, a Byzantine/BFT SMR service aims to satisfy two fundamental properties~\cite{pbft,berger2021making}:

\begin{enumerate}[leftmargin=20pt]
\item \emph{SMR Safety:} It behaves as a centralized service executing atomic operations, one at a time (linearizability~\cite{herlihy1990linearizability}).
\item \emph{SMR Liveness:} All operations issued by correct clients eventually complete.
\end{enumerate}

A common way to satisfy SMR Safety is to employ a consensus protocol for executing all operations/transactions in total order in all replicas, creating a replicated \emph{decision log} abstraction where every log position $i$ contains at most one decided operation (or batch of operations, as in blockchains).

There are a variety of Byzantine consensus protocols proposed in the literature for implementing BFT SMR.
In this paper, we are mostly interested in the ones that provide optimal resiliency ($t < \frac{n}{3}$) and best-case latency ($3$ communication steps~\cite{goodcaselatency}), such as PBFT~\cite{pbft} (described next).

Alternatives, such as speculative protocols like Zyzzyva~\cite{kotla2007zyzzyva} and ``fast'' consensus variants~\cite{martin2006fast,kuznetsov2021revisiting}, do not perform satisfactorily in geo-replicated settings. 
The main reason for this inadequacy lies in their network environment requirements and quorum formation rules. 
For instance, \emph{speculative execution}, as used in Zyzzyva, demands a predictable and stable network environment, which is uncommon in geo-distributed deployments.
Specifically, Zyzzyva's performance degrades because it necessitates responses from \emph{all} replicas within a \emph{strictly configured time window} to complete a request in a single phase~\cite{singh2008bft}.
On the other hand, fast Byzantine protocols run consensus in two communication steps but use proportionally \emph{larger} quorums, e.g., $4t-1$ out of $n=5t-1$~\cite{kuznetsov2021revisiting}. 
While theoretically faster in homogeneous networks, this approach also results in increased latency in geo-replicated settings~\cite{junqueira2007classic}.

\subsubsection*{PBFT}
The Practical Byzantine Fault Tolerance (PBFT) SMR algorithm~\cite{pbft} is considered the first practical method for implementing BFT services.
PBFT is optimal in terms of resilience and best-case latency, 
 ensuring safety under asynchrony and requiring a very weak form of synchrony for liveness.
PBFT orders requests by relying upon a stable leader that assigns sequence numbers to request batches.
If the leader is correct and the system is sufficiently synchronous, PBFT executes only its normal case operation. 
This pattern represents a \emph{Byzantine agreement/consensus instance} and consists of the leader proposing a batch of operations to all replicas (\textsc{Pre-Prepare}), followed by two phases of all-to-all message exchanges (\textsc{Prepare} and \textsc{Commit}), in which replicas use quorums to commit/decide the messages with a given sequence number despite Byzantine failures.
These quorums are sufficiently large to guarantee that any intersection of two quorums $Q$ and $Q'$ contains at least one correct replica, i.e., $|Q \cap Q' | \geq t + 1 $.
If the protocol stalls (e.g., the leader is faulty), a \emph{view change} sub-protocol is triggered when $t + 1$ replicas suspect the leader.
During a view change (alias leader change or synchronization phase~\cite{bftsmart}), the newly-elected leader collects the current status from a quorum of replicas and takes consistent decisions for pending requests.

\subsection{Weighted Quorums in BFT Replication}
\label{sec:wheat}

\wheat~\cite{wheat} improves PBFT-like SMR for geographically dispersed deployments by reducing client latency using $\Delta$ additional replicas, which does not affect the resilience threshold, i.e., $n = 3t + 1 + \Delta$.
Instead of the egalitarian Byzantine majorities of replicas used in most BFT works (e.g., \cite{pbft,hotstuff19,sui2022marlin}), \wheat{} uses weighted replication, which
allows for proportionally smaller quorums, achieved by selecting a well-connected clique of replicas with low-latency connections.
This approach maintains system availability, as votes from low-weight replicas are still used in case of failures.

BFT systems typically probe a Byzantine dissemination quorum containing $\lceil \frac{n+t+1}{2} \rceil$ replicas~\cite{malkhi1998byzantine}, as shown in Figure~\ref{fig:WHEAT_egalitarian}.
With $n=21$ replicas, quorums of size $14$ intersect in at least $t+1$ replicas (for $t=6$).
Achieving the \emph{same intersection property with smaller quorums} is possible by leveraging replicas with more voting power (\emph{weight}), as shown in Figure~\ref{fig:WHEAT_weighted}.
A fast quorum with $13$ replicas (voting weight $17$) is smaller than an egalitarian quorum while preserving intersection due to including all replicas with high voting power.
Consider the scenario where the fastest, geographically closest replicas constitute that quorum. 
These replicas can progress the voting phases of consensus more swiftly, as they need to wait less time for vote collection. 
This acceleration in \wheat{} also decreases the overall latency of BFT SMR~\cite{wheat}.

Unlike traditional protocols like PBFT that wait for responses from a strict number of replicas, \wheat{} waits for a sum of votes.
\wheat{} uses a bimodal scheme where the $2t$ best-connected replicas have a voting power of $V_{max} = 1+\frac{\Delta}{t}$ while the remaining replicas have a voting power of $1$.
As a result, the number of votes required for a quorum is $Q_v = 2t V_{max} + 1$.
This approach ensures the ability to form quorums even if $t$ best-connected replicas fail.
In this scheme, all quorums contain between $2t+1$ and $n-t$ replicas.
The size of the \emph{smallest} quorum only depends on the chosen threshold $t$, not on the actual size of the system $n$.
For instance, in Figure~\ref{fig:WHEAT_weighted_fast} a fast quorum comprises $7$ replicas for $t=3$.

\subsubsection*{AWARE}
Distributing voting weights is difficult due to the complexity of determining the optimal weight configuration for given network characteristics. 
\aware~\cite{aware} addresses this challenge by enabling geo-replicated state machines to self-optimize dynamically with automated weight tuning and leader placement,  supporting the emergence of fast quorums in the system.

In \aware, each replica monitors its latencies to other replicas by measuring the response times to protocol messages. 
These latency measurements are then disseminated by each replica with total order, and finally, all correct replicas consistently update a latency matrix~\cite{aware}.
This matrix is used as an input to a deterministic prediction model, which computes the expected consensus latency by simulating the protocol run for several configurations of weight distributions and leader locations, thus finding the optimal system configuration.

\aware{} automatically chooses the fastest configuration given a fixed resilience threshold $t$.
When the latency matrix is updated and changes are detected, the system automatically reconfigures the weight distribution and/or leader location to better suit the current network conditions.

\section{System Model and Design} \label{sect:system-model}
\subsubsection*{System Model}
This work employs the same system model used in BFT protocols such as HotStuff~\cite{hotstuff19} and PBFT~\cite{pbft}.
%
Ensuring liveness requires a weak \emph{partial synchrony}~\cite{dwork1988consensus} where the system may initially behave asynchronously and, after an \emph{unknown} GST (\emph{Global Stabilization Time}), some upper bound holds for all message transmission delays.
%
We consider an adaptive adversary capable of corrupting up to $t < n/3$ Byzantine replicas and an unbounded number of Byzantine clients.
Byzantine entities can behave arbitrarily and collude under the control of the adversary.

\subsubsection*{System Design}
\ourProtocol{} is a self-optimizing protocol transformation that adapts the resilience threshold of a BFT protocol and tunes replica voting weights to enable the emergence of smaller quorums for low-latency transaction ordering.
Our approach seeks to balance between maintaining maximum resilience and continuously striving for faster consensus execution by optimizing the system for the common case with few or no failures.
Specifically, we aim to design a fast BFT SMR approach for wide-area deployments that satisfies the standard SMR safety and liveness (see \S\ref{sec:state_machine_replication}) for the optimal resilience bound \mbox{$t=\lfloor \frac{n-1}{3} \rfloor$} and can tune itself to achieve fast commit latency when there is a stable, correct leader and no more than \mbox{$t_\mathit{fast} = \lceil \frac{t}{2} \rceil$} faulty replicas.

\begin{figure}[!t]
    \centering
    \includegraphics[width=\columnwidth]{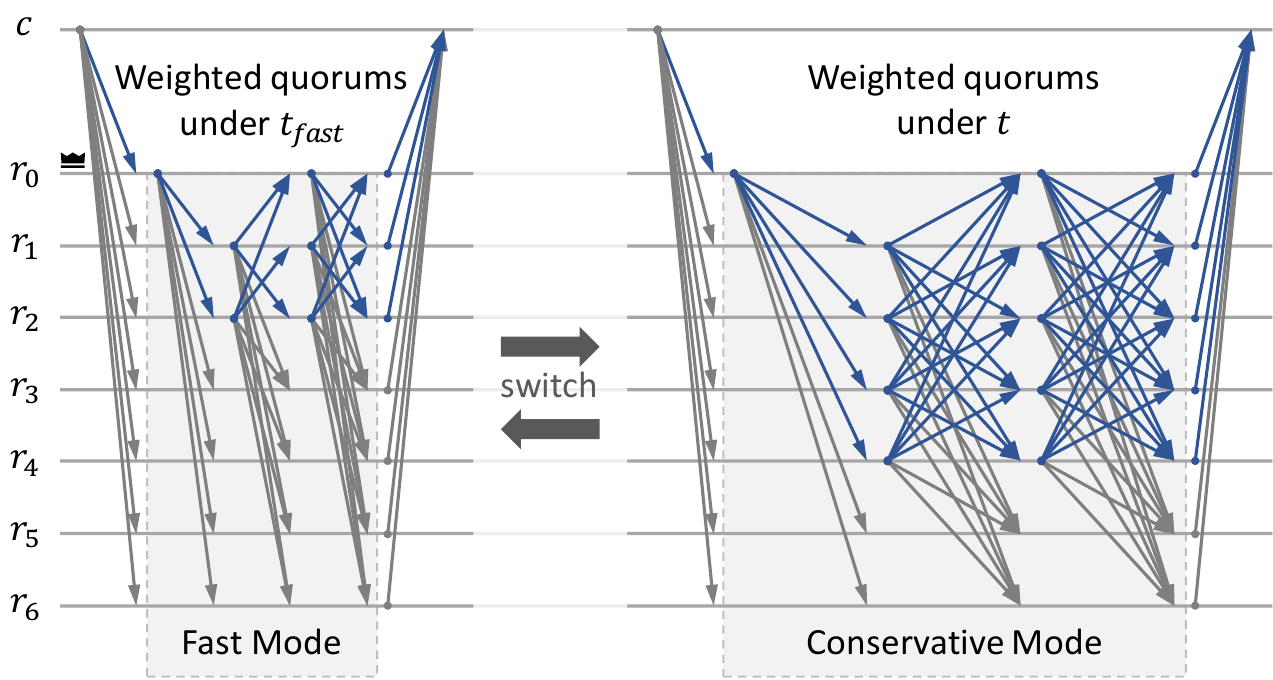}
    \caption{\ourProtocol{} two modes of operation ($t = 2$, $t_\mathit{fast} = 1$).}
    \label{fig:intro:reconfigurations} 
\end{figure}

This approach is implemented using two modes of operation (see Figure~\ref{fig:intro:reconfigurations}), as done in the past for improving performance~\cite{aublin2015next} and resource efficiency~\cite{distler2015resource}.
The system starts in \emph{conservative mode} by running instances of a quorum-based consensus protocol tolerating $t$ failures.
Periodically, after a number of consensus instances are decided, the system attempts to switch to the optimistic \emph{fast mode} that tolerates only $t_\mathit{fast}$ failures.
While the leader is correct and the number of actual failures $f$ does not surpass $t_\mathit{fast}$, \ourProtocol{} stays in this configuration and uses smaller quorums to accelerate consensus.
If latency improvements do not match expectations, the leader is suspected to be faulty, or correct replicas detect equivocations, \ourProtocol{} switches back to the conservative mode.


\subsubsection*{Building Blocks}
The weighted replication scheme introduced by \wheat{}~\cite{wheat} distributes voting power to enable the existence of small quorums, ensuring that all quorums intersect in at least one correct replica (consistency) and that there is always at least one quorum available (availability).
These requirements ensure that there is at least one minimal quorum with $2t_\mathit{fast}+1$ replicas, regardless of~$n$. 

Another hurdle when employing weighted replication is how to (re)assign the weights in accordance with the current system conditions.
In our case, we must select the $2t_\mathit{fast}$ replicas that will receive maximal voting power, so these can comprise the compact quorums.
We resort to the latency measurement and weight reassignment scheme of \aware~\cite{aware} to (re)assign weights (select the $2t_\mathit{fast}$ replicas that will receive maximal voting power) and focus instead on the problem of running the protocol optimistically, tolerating a few failures.
We remark, however, that the influence of malicious replicas on these measurements is bounded by \aware's sanitization strategy and the verification of a replica reported values using its geo-location, as done in secure location services~\cite{kohls2022verloc}.

\subsubsection*{Open Design Challenges}
Even while building upon the features of \aware~\cite{aware}, which allow us to use small quorums with reconfigurable weights based on the observed latency between replicas, the design of \ourProtocol{} encompasses several non-trivial challenges.
First and foremost, we need mechanisms to detect and diagnose the system when there are more than $f > t_\mathit{fast}$ failures.
Second, we need a robust reconfiguration mechanism to safely abort the fast mode and switch to the conservative one in such situations.
Finally, the client-replica contract must ensure linearizability in our dual fault-threshold approach.

\section{\ourProtocol: Threshold-Adaptive  BFT} 
\label{sec:ftlconsensus}

In this section, we describe \ourProtocol{}, a self-optimizing protocol transformation that adapts the resilience threshold of a BFT protocol to enable the emergence of smaller quorums for faster transaction ordering.
The consolidated algorithm of \ourProtocol{} is presented in Figure~\ref{fig:ftl-description}.
The algorithm starts with a client submitting a request $o$ to all replicas (\clientRule{\footnotesize C1}).
By \textbf{Rule}  (\serverRule{\footnotesize S1}), replicas start a \textit{timer} for each request, and the leader creates a batch of unordered requests and proposes it through the normal case operation of the underlying base protocol \textsc{Aware} or  \textsc{Aware}$\star$ if the system runs in $\mathit{fast}$ mode.
The rest of the algorithm summarizes all the required extensions on \aware{} to accommodate the novel mechanisms of \ourProtocol{}.
We describe these mechanisms in the context of the challenge they address in our protocol design.

\subsection{Challenge \#1: Dealing with $f > t_{fast}$ Failures}

A key challenge in devising \ourProtocol{} is ensuring safety and liveness when the system runs in fast mode and $f > t_\mathit{fast}$.

\begin{figure}[t]
    \centering

\begin{tcolorbox}[size=small,colback=lb4!20!white,colframe=db4,title= \textbf{\small Client}]
\small


\clientRule{\small C1} \textbf{Invocation:} 
Send $\langle \textsc{request}, o \rangle$ to all replicas.
 

\clientRule{\small C2} \textbf{Finalization:} Accept a result $\mathit{res}$ for $\mathit{o}$ if received a set of matching replies $rep=\{\langle \textsc{reply}, h(o), \mathit{fast}, \mathit{res} \rangle \}$ such that either:
\begin{enumerate}[leftmargin=*]
 \item $\mathit{fast} \land |rep| \geq n-t_\mathit{fast}-1$  OR
 \item $\neg \mathit{fast} \land  \sum_{r \in rep} weight(r) \geq 2t \cdot V_\mathit{max}+1$.
\end{enumerate}
In case of a timeout, keep re-sending the request and inspecting the decision log of replicas until one of the conditions above is satisfied.
 
 
\clientRule{\small C3} \textbf{Panic:} Broadcast $\langle \textsc{panic}, o, rep \rangle$ to the replicas if
replies 
$rep =\{ \langle \textsc{reply}, h(o), \texttt{true},* \rangle\}$ contains diverging results for operation $o$.
\end{tcolorbox}
\begin{tcolorbox}[size=small,colback=lb4!20!white,colframe=db4,title=\textbf{\small Replica}]
\small
\vskip 1pt
  \textbf{State}
  \vskip 3pt  
\setlength\tabcolsep{4.5pt}
\begin{tabularx}{\columnwidth}{lllX}
$\mathit{fast}$ & mode of operation         & boolean         & $\texttt{false}$      \\
$chkp$          & last stable checkpoint    & bytes           & \texttt{null} \\
\end{tabularx}%
   \vskip 4pt
 \textbf{Building Blocks} 
  \vskip 1pt 
\begin{tabular}{llll}
\small
\textsc{aware} & \aware{} SMR protocol (conservative)    \\
\textsc{aware}$\star$ &  normal case operation of \aware{} in fast mode    \\
\textsc{audit} & lightweight forensics procedure of Figure~\ref{fig:forensics}   \\  
\end{tabular}%
\tcblower
\small


\serverRule{\small S1} \textbf{Request Processing:} Start a $\mathit{timer}$ for each received client request. 
If leader, create a batch of requests and propose it using \textsc{aware}$\star$, if $\mathit{fast}$, or \textsc{aware}, otherwise.
 

\serverRule{\small S2} \textbf{Periodic Checkpoint:} When a snapshot of the service state $chkp'$ is created after processing consensus $j$, broadcast signed message $\langle \textsc{checkpoint}, h(chkp'), j \rangle$.
If $n-t$ matching checkpoint hashes $h(chkp')$ for $j$ are received, update the last stable checkpoint $chkp$ to $chkp'$.
If there are no $n-t$ matching checkpoints, run \textsc{audit}.

 
\serverRule{\small S3} \textbf{Client Panic:} If $\mathit{fast}$ and received a message $\langle \textsc{panic}, o, rep \rangle $ with diverging signed replies for $o$ in $\mathit{rep}$ from a client, run \textsc{audit}.

 
\serverRule{\small S4} \textbf{Abort:} If $\mathit{fast}$: Broadcast a \textsc{view-change} message if one condition applies:
(1) a request $\mathit{timer}$ expires, 
(2) a message $\langle \textsc{poc}, poc \rangle$ with a valid PoC is received from some replica, or 
(3) upon \emph{consensus latency disappointment}---see \S\ref{subsec:performancedegradation}.


\serverRule{\small S5} \textbf{Switch:} After deciding $\theta$ consensus instances in a row using \textsc{aware}, set $\mathit{fast}$ to $\texttt{true}$ (the optimization interval $\theta$ is inherited from \aware).

 
\serverRule{\small S6} \textbf{Synch. Phase:}
Upon receiving $t+1$ matching \textsc{view-change} messages, use \textsc{aware}' synch. phase to replace the current leader and synchronize the decision log.
If $\mathit{fast}$, set $\mathit{fast}$ to $\texttt{false}$ and run \textsc{audit} if no $\langle \textsc{poc}, poc\rangle$ message has been received so far. 
\begin{enumerate}[leftmargin=*]
    
    \item When waiting for \textsc{new-view} messages, the next leader checks for PoCs and ignores messages from equivocating replicas.
    When consolidating operations for each position of the decision log, the new leader picks the most commonly reported prepared value.
    \item Upon a PoC is produced or received during the synch. phase, all replicas roll back to $chkp$ and use the decisions (with proofs) obtained from the new leader to re-execute decided operations.
    The new leader proposes $\langle \textsc{reconfigure}, \mathit{culprits}, poc \rangle$ using \textsc{aware}.
    \item Upon deciding $\langle \textsc{reconfigure}, \mathit{culprits}, poc \rangle$, a replica verifies the $poc$ using \textsc{audit}, and removes the $\mathit{culprits}$ from the system.
\end{enumerate}
\end{tcolorbox}
\vspace{-1mm}
\caption{A summary of \ourProtocol{}.}
\label{fig:ftl-description}
\vspace{-2mm}
\end{figure}
\begin{figure}[t]
\centering
\begin{tcolorbox}[size=small,colback=lb4!20!white,colframe=db4,title= \textbf{\small Replica}]
\small
\auditRule{\small F1} \textbf{Find evidence:} Let $S$ and $S'$ be the two sets of replicas with diverging checkpoint digests.
The auditor tries to collect signed lists of decision proofs from consensus instances $i-k+1$ to $i$ from at least one of the replicas of each of these two sets.
\vskip 3pt  

\auditRule{\small F2}  \textbf{Produce PoC:} When such logs are obtained, the auditor checks the logs to find the first consensus instance with diverging decisions.
Once such an instance is found, the auditor checks the proofs of decisions.
\begin{enumerate}
    \item If any of the proofs of decision is invalid, the log signed by the replica that provided it is a PoC for the replica.
    \item If both proofs are valid, the auditor finds at least $t_\mathit{fast}+1$ malicious replicas that provided signed \textsc{Accept} messages for both decisions.
    These two conflicting proofs are the PoC (proof-of-culpability) for the 
    malicious replicas. 

\end{enumerate}
\auditRule{\small F3} \textbf{Blame culprits:} If $poc\neq\emptyset$: Broadast a $\langle\textsc{poc}, poc \rangle$ message.
\end{tcolorbox}
\vspace{-1mm}
\caption{Lightweight forensics procedure.}
\label{fig:forensics}
\vspace{-2mm}
\end{figure}

\subsubsection{Safety}
\label{subsec:safety}

When running in fast mode, the adversary can control more than $t_\mathit{fast}$ replicas and cause equivocations in the system.
This situation might lead correct replicas to decide different transaction batches in a consensus instance since fast mode's smaller quorums are not guaranteed to overlap in one or more correct replicas.

\paragraph{Periodic Checkpoint} To handle this scenario, the system must detect if the actual number of faults ($f$) surpasses the resilience threshold of the fast mode ($t_\mathit{fast}$) and, if necessary, revert the system to the conservative mode.
We need to check the state of the replicas through periodic \emph{checkpoint messages} (as in PBFT~\cite{pbft}) to ensure they are consistent and detect faults.
By \textbf{Rule}~\serverRule{\small S2}, on every $k$ completed consensus instances, each replica takes a snapshot of the service state and broadcasts to all replicas a signed message with the digest of this snapshot $h$ and the highest consensus instance $i$ that affected it.
Each replica waits for $n-t$ matching checkpoint hashes for the same consensus instance to define the checkpoint as \emph{stable}.
During this process, if a correct replica, which we call \emph{the auditor}, detects non-matching checkpoints, it runs the lightweight forensics procedure of Figure~\ref{fig:forensics} to identify and obtain a non-repudiable \emph{Proof-of-Culpability}~(PoC) for the protocol violators.

This protocol can identify equivocating replicas in the system if there are no more than $2t_{fast}$ faulty replicas. 
In fact, Sheng et al. show that it is impossible to identify misbehaving replicas if there are more than $2t$ Byzantine replicas in a system tolerating $t$ failures~\cite{bftforensics}.
This limitation leads \ourProtocol{} to use $t_\mathit{fast} = \lceil \frac{t}{2} \rceil$.

If the auditor receives $n-t$ matching checkpoint hashes during the procedure execution, it stops the forensics procedure.
This action prevents faulty replicas from blocking correct replicas in forensics procedures since a faulty replica can send a non-matching checkpoint but never send its corresponding log.
%
After concluding the lightweight forensics procedure, if a PoC is produced for one or more replicas, the auditing replica broadcasts this PoC to all replicas, forcing the system to switch to the conservative mode and expel the misbehaving replicas (described in~\S\ref{subsec:reconfiguration}).

\paragraph{Client Panic}
The lightweight forensics protocol also triggers if a client detects non-matching signed replies for a request (\textbf{Rule}~\clientRule{\small C3}).
When this happens, the client sends a \textsc{Panic} message with conflicting replies to the replicas.
By \textbf{Rule}~(\serverRule{\small S3}), a replica that receives a \textsc{Panic} message with correctly signed conflicting replies starts the lightweight forensics protocol, but fetches logs from the last checkpoint until the consensus instance that decided the problematic request.

\subsubsection{Liveness}
\label{subsec:liveness}

Besides equivocations, $f>t_\mathit{fast}$ replicas controlled by an adversary can stay silent and negatively affect the liveness of the system.
In such situations, there will be fewer than $n-t_\mathit{fast}$ correct replicas in the system, violating a key liveness assumption of the consensus protocol designed for no more than $t_\mathit{fast}$ failures.
This can lead to two unfavorable situations.
First, client requests might not be ordered, triggering a timeout and initiating a view change.
As explained in the next section, this sub-protocol reverts the system to the conservative mode that tolerates up to $t$ failures.

The second situation is more complicated: the request might be ordered, but faulty replicas might not send replies to the client, preventing it from consolidating the request result.
In this case, the client could send a \emph{panic} message to the replicas asking them to switch to the conservative mode, which might trigger a leader change to switch the system's mode.
However, this must be done carefully, as a malicious client could abuse this mechanism to prevent the system from operating in fast mode, a known weakness inherent to optimistic protocols~\cite{aublin2015next}.
This mechanism could make \ourProtocol{} optimization fragile, as a single malicious client can undermine latency improvements. 
Thus, we propose an alternative approach for dealing with this situation.

\paragraph{Finalization}
By \textbf{Rule}~\clientRule{\small C2}, if the client does not receive the required number of replies, it periodically checks the decision log until the next checkpoint to see where its operation appears in the finalized decision log.
This procedure is similar to how blockchain clients inspect the blockchain until their requests are included in a block several blocks away from the blockchain head.
This approach ensures that clients can benefit from \ourProtocol{}'s low latency as long as there are no more than $t_\mathit{fast}$ faulty replicas in the system. 

\subsubsection{Performance Degradation}
\label{subsec:performancedegradation}

To ensure \ourProtocol{} does not lead to performance degradation when compared to the conservative mode, all replicas periodically monitor their observed performance and compare it with expectations they have on the conservative mode.
Replicas retrieve their expectations from \aware's underlying latency prediction model~\cite{aware}. 
Using this model, replicas can predict their consensus latency for the conservative mode using the network latency map and set their \emph{consensus latency expectation threshold}.\footnote{
In \aware, replicas run consensus to create a uniform view of latency measurements, which are recorded in the decision log. 
These can be accessed by each replica to deterministically compute an expectation value for consensus latency by feeding the measurements into a model that simulates the protocol run in conservative mode.
The obtained expectation value serves as an indication to determine if an enabled optimization actually accelerates consensus or not.
}
If replicas find that the latency they currently observe exceeds this threshold, they stop their execution and ask for a view change (\serverRule{\small S4}).
When $t+1$ replicas ask for a view change, the system switches to the conservative mode (see next section).
In the end, \ourProtocol{} only runs in fast mode if replicas observe the consensus latency to be lower than the expected latency in the conservative mode.

\subsection{Challenge \#2: Reconfiguration of the System}
\label{subsec:reconfiguration}

As explained before, \ourProtocol{} operates in two regimes: fast, in which smaller quorums are used and $t_\mathit{fast}$ failures are tolerated, and conservative, in which standard-size quorums are employed and $t$ failures are tolerated.

\paragraph{Switch}
The system starts in the conservative mode, and by \textbf{Rule}~\serverRule{\small S5} after finishing a predefined number of $\theta$ consecutive consensus instances, it switches to the fast mode.
Such reconfiguration is very simple because it is done deterministically at a certain point in the execution, i.e., after a certain consensus instance is decided.
At this point, it simply requires each replica to locally change the fault threshold to $t_\mathit{fast}$ and recalculate its quorum size before executing the next consensus instance.

\paragraph{Abort}
A replica stays in the fast configuration until the underlying algorithm view change is triggered by \textbf{Rule}~\serverRule{\small S4}.
This approach can be done deliberately due to either safety or liveness issues, as discussed in previous subsections.
In both cases, we require the participation of $t+1$ replicas to start the view change, which always runs considering threshold~$t$, not $t_\mathit{fast}$ (which might already be violated).

\paragraph{Synchronization Phase}
During the synchronization phase (\textit{view change} in PBFT parlance), the newly elected leader receives from $n-t$ replicas the log of the decided instances since the last checkpoint and verifies it for diverging decisions using the lightweight forensics procedure (Figure~\ref{fig:forensics}). \textbf{Rule}~\serverRule{\small S6} is applied in subsequent steps: 
\begin{enumerate}
\item If multiple decisions for the same slot exist, the new leader selects the most commonly reported one to consolidate the decision log.
\item The first transaction of the newly elected leader's regency, after repairing the system to a single transaction history, is a reconfiguration request~\cite{bftsmart}. This request aims to remove the Byzantine replicas involved in the equivocation and contains the PoC generated during the forensics procedure.
Correct replicas remove these compromised replicas from the system after processing reconfiguration requests with valid PoCs.
\item Further, if the replica was subject to equivocation and the new leader decided differently from it, it might need to roll back its state to a previous \emph{stable checkpoint} (see \S\ref{subsec:safety}), reapplying the correct transaction history as defined by the new leader on this state.
\end{enumerate}

\begin{figure*}[t]
    \centering
    \begin{subfigure}{\columnwidth}
        \centering
    \includegraphics[width=0.8\columnwidth]{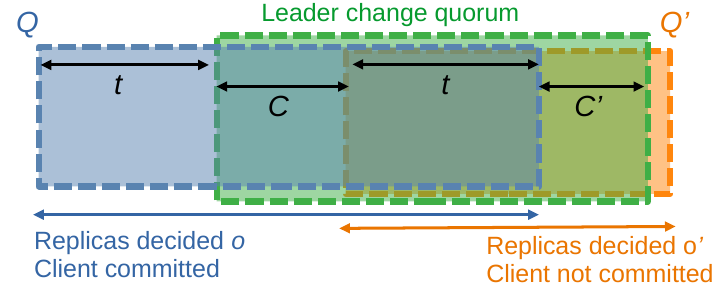}
    \caption{Using only status information from replicas.}
    \label{fig:quorum-reasoning-status}
    \end{subfigure}
    \begin{subfigure}{\columnwidth}
        \centering
    \includegraphics[width=0.8\columnwidth]{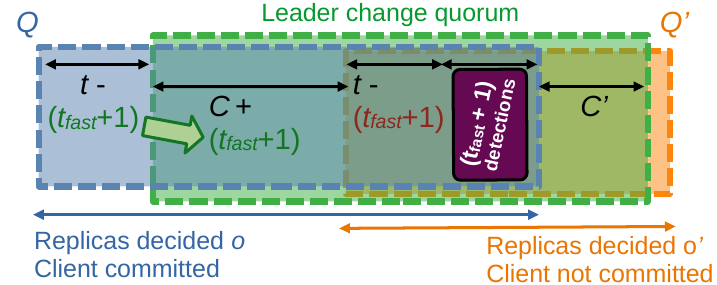}
    \caption{Using status and audit information from replicas.}
     \label{fig:quorum-reasoning-audit}
    \end{subfigure}
    \caption{Quorum reasoning in \ourProtocol.}
    \label{fig:quorum-reasoning}
\end{figure*}

\subsection{Challenge \#3: Ensuring Linearizability}
\label{sec:clientguarantees}

In typical BFT SMR systems, a client waits for $t+1$ matching replies to ensure the replicated system perfectly emulates a centralized server, satisfying linearizability~\cite{herlihy1990linearizability}.
This quorum size becomes $\lceil \frac{n+t+1}{2} \rceil$ if one wants to avoid running a consensus for read-only operations~\cite{pbft,berger2021making}.
These quorum sizes are still valid in \ourProtocol{} while in conservative mode; however, when the system is in fast mode, the existence of equivocations and the possibility of divergent decisions (that will be later detected and punished) requires revisiting the number of matching replies expected by clients in \textbf{Rule}~\clientRule{\small C2}.

Figure~\ref{fig:quorum-reasoning} illustrates the scenario where two clients received replies for operations $o$ and $o'$ from two different quorums $Q$~and~$Q'$, respectively, for consensus instance~$i$ (ignore the leader change quorum for now in the Figure~\ref{fig:quorum-reasoning-status}).
Even if $t$ malicious replicas are present in the intersection of the quorums, a client can assume that its request has been committed and will not be rolled back by waiting for $n-t$ matching replies. 
This holds because the intersection $(n-t) + (n-t) - n > t$ when $n > 3t$. 
It means that two quorums with $n-t$ replicas intersect in more than $t$ of them, ensuring the presence of at least one correct replica in this intersection.
By waiting for $n-t$ matching replies, the responses accepted by clients will never be rolled back, even with divergent decisions for consensus instance~$i$, \emph{as long as there are no leader changes}.

Now, consider the same scenario in which the replies for~$o$ were deemed final by the client, but there was a leader change, and the new leader needs to define the result of consensus instance $i$.
In this scenario, the elected leader waits for $n-t$ replicas to inform their status as indicated in the leader change quorum of Figure~\ref{fig:quorum-reasoning-status}, receiving replies from every replica but $t$ slow replicas that decided operation~$o$.

In this setting, the decision of $o$ will be preserved as long as such value is the majority value among the ones informed by replicas, i.e., $C > C' + t$ in the figure.
Considering $n = 2t +C +C'$ and $|Q| = C+2t$ (both directly from the highlighted variables in the figure), we can reach that $C > \frac{n}{3}$, leading to $|Q| = n$.
Therefore, \emph{waiting for $n-t$ matching replies is insufficient to ensure a value will never be rolled back during a leader change that switches the system to the conservative mode.}
The only quorum big enough to ensure this is waiting for matching replies from \emph{all replicas}.

Fortunately, integrating a continuous lightweight BFT forensics procedure in the view change sub-protocol (view change in PBFT or synchronization phase in \bftsmart{}) enables the use of smaller quorums.
More specifically, we observe that to produce equivocations that lead some correct replicas to decide $o$ and $o'$, and later force a committed value to be rolled back, the $t_\mathit{fast}+1$ equivocators must participate in the three quorums (for~$o$, $o'$, and leader change).
Therefore, if the new leader executes the forensics protocol during the leader change, it detects $t_\mathit{fast}+1$ equivocators.
Consequently, it can discard the contributions of these malicious replicas and wait for messages from $t_\mathit{fast}+1$ additional replicas.
This situation is illustrated in Figure~\ref{fig:quorum-reasoning-audit}.


In this scenario, instead of assuming $C > C' + t$, we have 
\begin{equation}
C + (t_\mathit{fast}+1) > C' + t - (t_\mathit{fast}+1)
\end{equation}
By developing this in inequality like before, we find that \emph{by waiting for \mbox{$n-t_\mathit{fast}-1$} matching replies in fast mode, a client knows the result of its operation is finalized, ensuring the durability and linearizability of the replicated service}, allowing us to prove the following theorem in the appendix of this paper:

\vspace{3mm}
\noindent
\textbf{Theorem 1.}
\emph{If an operation $o$ is finalized in \mbox{$i$-th} position of the decision log, then no client observes an operation $o'\neq o$ in this position of the decision log.}
\vspace{1mm}

\paragraph{Detailed derivation of $Q = n-t_\mathit{fast}-1$:}
Considering $n=2t+C+C' > 3t$ (from Figure~\ref{fig:quorum-reasoning-status}), it follows that $C+C' > t$ and thus $C' > t  - C$.
Replacing $C'$ with this value in inequality (1) yields:
\begin{equation*}
	C + (t_{fast}+1) > \overbrace{t - C}^{\mathclap{C'}} + t - (t_\mathit{fast}+1)    
\end{equation*}
\begin{equation*} 
	2C  >  2t - 2(t_\mathit{fast}+1)   
\end{equation*}
\begin{equation}
	C  >  t - t_\mathit{fast} - 1   
\end{equation}
Now, we calculate the response quorum $Q  =  2t + C $ (also from Figure~\ref{fig:quorum-reasoning-status}) using the inequality of (2):
\begin{equation*} 
	Q  >  2t + t - t_\mathit{fast} - 1
\end{equation*}
\begin{equation*} 
	Q  >  3t - t_{fast} - 1    
\end{equation*}
Which can be generalized to $Q = n-t_\mathit{fast}-1$.

\section{Implementation and Optimizations}
\label{sect:implementation}

\paragraph{Implementation}
\ourProtocol{} was implemented on top of the \aware{} prototype~\cite{aware}, which is based on BFT-SMaRt.
We stress that all mechanisms employed in \aware{} (e.g., latency measurement and weights reassignment) could be implemented in any quorum-based SMR protocol.
This implementation uses TLS to secure all communication channels and the elliptic curve digital signature algorithm (ECDSA) and SHA256 for signatures and hashes, respectively.
Most of our modifications are related to the switching between two modes of operation (with different resilience thresholds) and implementing BFT forensics.

\begin{figure}[t]
    \centering
    \includegraphics[width=0.9\columnwidth]{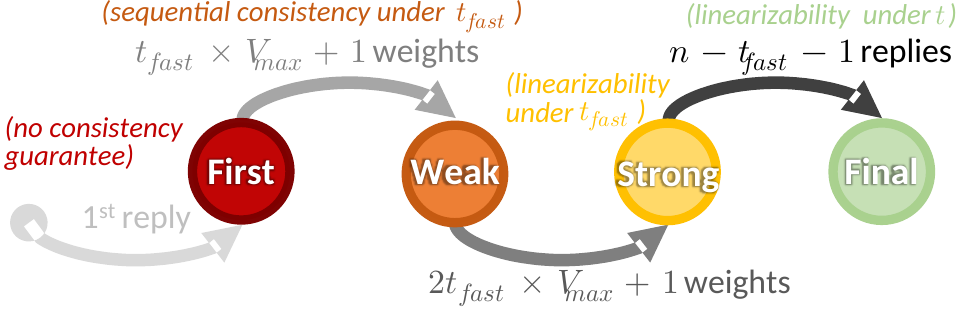}
    \caption{Incremental consistency levels that can be accessed through the Byzantine correctable programming interface.}
    \label{fig:correctable-programming}
\end{figure}

\subsection{Improving Latency with Speculation}

Operating in fast mode requires clients to collect \mbox{$n-t_\mathit{fast}-1$} matching replies to preserve linearizability.
This finalization quorum is considerably bigger than the $t+1$ replies typically required in BFT SMR and is expected to negatively impact clients' observed latencies in fast mode. 
\ourProtocol{} can lower the latency observed by clients further using client-side speculation.
For this purpose, we implemented correctables~\cite{guerraoui2016incremental} in the client shim of our BFT protocol.
A correctable is a programming abstraction that allows a client application to work with incremental consistency guarantees and accelerates the application by allowing it to speculate with intermediate results.
For example, this enables fast confirmation for low-value transactions such as time-sensitive micropayments even before linearizability under $t$ failures is ensured.
The state of a correctable can be updated multiple times, depending on the replies received by the client, strengthening the consistency guarantee each time until it reaches the final state, which corresponds to the strongest consistency guarantee.
\ourProtocol{}'s \textit{Byzantine correctables} follow two principles:
\begin{enumerate}
    \item Ensure the same safety guarantee as traditional BFT SMR: the final consistency guarantee must satisfy linearizability under the resilience threshold~$t$.
    \item Less safe consistency guarantees may relax either the assumptions on the number of Byzantine replicas or trade linearizability for a weaker consistency model.
\end{enumerate}

We define incremental consistency levels for \ourProtocol{} as follows (see Figure~\ref{fig:correctable-programming}).
\emph{First} is the speculative result a client can access as soon as the first response arrives, which does not provide any correctness guarantee.
\emph{Weak} demands replies from replicas totaling $t_\mathit{fast} \times V_\mathit{max} + 1$ votes, being $V_\mathit{max}$ the maximum weight assigned to a replica, and thus must have been confirmed by at least one correct replica if $f \leq t_\mathit{fast}$.
This result can be stale, satisfying only sequential consistency~\cite{attiya1994seqconsvslin} under $t_\mathit{fast}$ failures.
\emph{Strong} demands $Q_v=2t_\mathit{fast} \times V_\mathit{max} + 1$ votes and satisfies linearizability if $f \leq t_\mathit{fast}$.
Lastly, the \emph{Final} level satisfies linearizability under $t$ (just like any typical SMR with the read-only optimization enabled) by waiting for $n-t_\mathit{fast}-1$ replies, as explained in the previous section.
Since classical SMR preserves linearizability~\cite{pbft}, only \emph{Strong} and \emph{Final} give the typical safety guarantee for their respective resilience thresholds.

\section{Evaluation}
\label{sec:evaluation}

We evaluate the latency of the \ourProtocol{} prototype on the AWS cloud as well as a simulated network of $51$ replicas based on real data from the internet.
Parts of our experiments were conducted in our local data center using high-fidelity tools for network emulation and simulation~\cite{gouveia2020kollaps,jansen2022co}.
We validated the fidelity of these emulated/simulated setups with additional experiments reported in the appendix of this paper.
Our experiments focus on measuring latency, which is fundamentally limited by the distance between nodes and quorum formation rules in WANs.


\begin{figure*}[!t]
\centering
    \begin{subfigure}{0.19\textwidth}
     \centering
         \begin{tikzpicture} 
    \begin{axis}[ 
     font= \small,
     ylabel={Latency [ms]}, 
     xlabel={Protocol}, 
     xticklabels=\none,
     xticklabel style={rotate=45},
     ybar=0pt,  
     bar width=8pt,
    height=4cm,
       ytick = {0,50,100,150,200,250, 300, 350, 400},
       xmin=-0.5,
       xmax=4.5,
        ymin=0,
        ymax=400,
    ymajorgrids=true,
    yminorgrids=true,
    minor grid style={dashed,gray!10},
legend style={at={(1, -0.09)},
    legend columns = 1,
    legend cell align=left
    }
    ]

\addplot[db1, fill=lb1] coordinates {
    (1, 368)
};   \addlegendentry{\bftsmart/PBFT};

\addplot[db3, fill=lb3] coordinates {
(2, 231)
};   \addlegendentry{AWARE};

\addplot[db4, fill=lb4]  coordinates {
(3, 103)
};   \addlegendentry{\ourProtocol};

\draw[black, thick] (axis cs:-0.5, 107.928) -- (axis cs: 5, 107.928);
\draw[gray, thick] (axis cs:-0.5, 161.086567164179) -- (axis cs: 5, 161.086567164179);

\end{axis} 
\path (current bounding box.north) ++ (0,0.4cm);
\end{tikzpicture} 
      \caption{Consensus latency.}
    \label{fig:aws21-consensus-latency-B}
\end{subfigure}
  \begin{subfigure}{0.8\textwidth}
      \centering
\centering
\begin{tikzpicture} 
    \begin{axis}[ 
     ylabel={Latency [ms]}, 
     xticklabels from table={data/eval-real-aws.txt}{region},    
     ybar=0.5pt,  
         font= \small,
     bar width=4pt,
    width=\linewidth, 
    height=4.2cm,
       xtick=data, 
            xticklabel style={rotate=10},
       ytick = {0,100,200,300,400,500,600, 700, 800},
        ymin=0,
        ymax=900,
    ymajorgrids=true,
    yminorgrids=true,
    minor grid style={dashed,gray!10},
    minor tick num=1,
    legend style={at={(1, 1.25)},
    legend columns = 6,
    legend cell align=left
    }
    ] 
      \addplot 
       [draw = db1, 
        fill = lb1]   
        table[ 
          x=regionNr, 
          y=bftsmart    
          ] 
      {data/eval-real-aws.txt}; 
      \addlegendentry{\bftsmart/PBFT}; 

          \addplot 
       [draw = db3, 
        fill = lb3]   
        table[ 
          x=regionNr, 
          y=aware    
          ] 
      {data/eval-real-aws.txt}; 
      \addlegendentry{AWARE \ \ \ \ \ourProtocol:}; 
   
            \addplot 
       [draw = dark-green, 
        fill = light-green,
        ] 
        table[ 
          x=regionNr, 
          y=final    
          ] 
      {data/eval-real-aws.txt}; 
       \addlegendentry{final}; 
       \addplot 
       [draw = dark-yellow, 
        fill = light-yellow,
        ]   
        table[ 
          x=regionNr, 
          y=strong    
          ] 
      {data/eval-real-aws.txt}; 
      \addlegendentry{strong};
      
      \addplot 
       [draw = dark-orange, 
        fill = light-orange,
        ]   
        table[ 
          x=regionNr, 
          y=weak    
          ] 
      {data/eval-real-aws.txt}; 
      \addlegendentry{weak};
      
      \addplot 
       [draw = dark-red, 
        fill = light-red,
        ]   
        table[ 
          x=regionNr, 
          y=none   
          ] 
      {data/eval-real-aws.txt}; 
      \addlegendentry{first};

\draw[black, very thick] (axis cs:1, 840) -- (axis cs: 1.5,840);
\node[right, black] at (axis cs:1.5 ,840) {{\small \bftsmart/PBFT at 1c}};
\draw[gray, very thick] (axis cs:3.5, 840) -- (axis cs: 4,840);
\node[right, gray] at (axis cs:4 ,840) {{at 0.67c}};

    \foreach \region \latency in {
1/218.7635415625,
2/231.2720105,
3/206.641676875,
4/194.887005625,
5/212.910737083333,
6/234.59386875,
7/205.239995
    } 
    {
     \edef\temp{\noexpand\draw[black, thick] (axis cs:\region - 0.26, \latency) -- (axis cs: \region + 0.26, \latency);}
     \temp
}

    \foreach \region \latency in {
1/326.512748600746,
2/345.182105223881,
3/308.420413246269,
4/290.876127798507,
5/317.777219527363,
6/350.14010261194,
7/306.328350746269
    } 
    {
     \edef\temp{\noexpand\draw[gray, thick] (axis cs:\region - 0.26, \latency) -- (axis cs: \region + 0.26, \latency);}
     \temp
}

    \end{axis}

    
\end{tikzpicture} 

\caption{End-to-end latencies observed by clients in protocol executions with BFT-SMaRt, AWARE, and \ourProtocol. Client results are averaged over all regions per continent.}
\label{fig:kollaps-aws-latencies}
  \end{subfigure}
  \caption{
  Achievable latency improvements for the $n=21$ AWS setup.}
  \label{fig:eval:aws21}
\end{figure*}

\begin{figure*}
    \centering
     \begin{tikzpicture}
    \begin{axis}[
width= \linewidth,
height=4.8cm,
font= \small,
    xlabel={Time [s]},
    ylabel={Latency [ms]},
    xmin=0, xmax=3300,
    ymin=0, ymax=1000,
    xtick={0, 300, 600, 900, 1200, 1500, 1800, 2100, 2400, 2700, 3000, 3300},
    xticklabel style={/pgf/number format/1000 sep={}},
    ytick={0, 100, 200, 300, 400, 500, 600, 700, 800, 900, 1000},
    yticklabel style={/pgf/number format/1000 sep={}},
    legend columns = 4,
    legend style={at={(0.15,0.92)},anchor=west, legend columns=4},
    legend cell align={left},
    ymajorgrids=true,
    xmajorgrids=true,
    grid style=dashed,
]

\addplot[
    color=light-red,
    mark=.,
    ]
    table [x=time,y=none] {data/runtime2.txt};  \addlegendentry{first}; 

\addplot[
    color=light-orange,
    mark=.,
    ]
    table [x=time,y=weak] {data/runtime2.txt}; \addlegendentry{weak}; 

    \addplot[
    color=light-yellow,
    mark=.,
    ]
    table [x=time,y=strong] {data/runtime2.txt}; \addlegendentry{strong};

        \addplot[
    color=light-green,
    mark=.,
    ]
    table [x=time,y=final] {data/runtime2.txt}; \addlegendentry{final};

 \draw (axis cs: 0, 800) rectangle (axis cs: 277, 1000)  node[pos=.5] {\scriptsize conservative};
 \draw (axis cs: 277, 400) rectangle (axis cs: 1814, 600)  node[pos=.5] {\scriptsize fast mode };
    \draw (axis cs: 1814, -50) rectangle (axis cs: 1846, 150)  node[pos=.5] {\scriptsize leader failure};
    \draw (axis cs: 1846, 800) rectangle (axis cs: 1888, 1000)  node[pos=.5] {\scriptsize abort};
    \draw (axis cs: 1888, 600) rectangle (axis cs: 2022, 800)  node[pos=.5] {\scriptsize switch};
    \draw (axis cs: 2022, 400) rectangle (axis cs: 3400, 600)  node[pos=.5] {\scriptsize fast mode };
    \draw (axis cs: 2605, -50) rectangle (axis cs: 3400, 150)  node[pos=.5] {\scriptsize malicious \textit{eu-west-2}};
      \draw (axis cs: 2827, 600) rectangle (axis cs: 3400, 800)  node[pos=.5] {\scriptsize after optimization};

\end{axis}
\end{tikzpicture} 
    \caption{Runtime behavior of \ourProtocol{} under induced failures.}
    \label{fig:ftl-runtime}
\end{figure*}

\subsection{AWS Data-centers}

To begin with, we investigate the potential performance gains of \ourProtocol, comparing it to \aware{} and \bftsmart{} as baselines.
Later, we reason about \ourProtocol's runtime behavior (particularly its adaptiveness) in the presence of failures or malicious replicas.

\subsubsection*{Setup}
We use \texttt{c5.xlarge} instances on the AWS cloud for deploying a client and a replica in each of the $n=21$ AWS regions (depicted in Figure~\ref{fig:aws21-WHEAT-quorums}).
All clients send $400$-bytes requests simultaneously and continuously to the replicas ($2000$ requests per client) until each client has finished its measurements. 
A client request arriving at the leader may wait until it gets included in a batch when there is currently a consensus instance running. 
We employ synchronous clients that block until a result is obtained and send the next request after randomly waiting for up to $1$s.
Finally, \emph{request latency} is the average end-to-end protocol latency computed by a client after finalizing all operations.

\subsubsection{\ourProtocol{} Acceleration}
\label{sect:eval:kollaps-speedup}

For a better exposition, we group the $21$ clients' results by the continent they are located in, reporting only their regional averages (see Figure~\ref{fig:eval:aws21}).
First, we observe that \ourProtocol{} significantly accelerates consensus, leading to a speedup of $3.57\times$ for reaching decisions\footnote{To put these results in perspective, the \ourProtocol{} latency (100ms) is more than $4\times$ faster than the results reported for a recent optimistic protocol in a similar network (see Table 1 in \cite{boltdumbo} and Figure 4 in \cite{babel2024mysticeti}).} (Figure~\ref{fig:aws21-consensus-latency-B}).
This result also surpasses the speedup of $2.29\times$, achievable if the speed of the links employed by \bftsmart/PBFT approaches the speed of light.\footnote{Which is impossible: in practice, it is accepted that the maximum speed internet links can reach is $0.67c$ \cite{cangialosi2015ting,kohls2022verloc}.}
Second, accelerating consensus decisions also leads to faster request latencies observed by clients worldwide (Figure~\ref{fig:kollaps-aws-latencies}). 
Averaged over all client regions, \ourProtocol{} leads to a speedup of $1.87\times$ over \bftsmart{} for clients' observed end-to-end request latencies with \textit{Final} consistency (\aware{} with the same resilience leads to $1.33\times$ only).

Our results also show that even higher speedups can be achieved by employing the incremental consistency levels of \ourProtocol{}' correctables.
The \emph{Strong} consistency level, which guarantees linearizability if $f < t_\mathit{fast}$, achieves a speedup of $2.38\times$, while the speculative levels \emph{Weak} and \emph{First} achieve speedups of $2.76\times$ and $2.90\times$, respectively (results are averaged over all regions).

\subsubsection{Runtime Behavior under Failures}
For this experiment, we create an emulation of the AWS network in our local cluster to be more flexible with the induction of events.
The emulated network uses latency statistics from \emph{cloudping},\footnote{\url{https://www.cloudping.co/grid}.} and, the Kollaps network emulator, which was validated for realistic WAN experimentation with \bftsmart{} and \wheat~\cite{gouveia2020kollaps}. 
We launch \ourProtocol{} in the same $n=21$ AWS regions to observe its runtime behavior during the system’s lifespan. 
Noticeably, clients’ request latencies show high variations, which are caused by a random waiting time of a request at the leader before getting included in the next batch, which takes a varying time depending on how shortly the request arrives before the next consensus can be started.
Moreover, we induce the following events to evaluate \ourProtocol’s reactions and plot the latency observed by a representative correct client in Figure~\ref{fig:ftl-runtime}.

\emph{Configuration switch.}
\ourProtocol{} starts in the conservative configuration, displaying a latency similar to a ``normal'' PBFT-like protocol.
Later, around time $277$s, the system switches to the fast configuration (such switches are attempted every $k=400$ consensus instances), leading to a significant latency improvement.
    
\emph{Silent leader.}
At the time $1814$s, the leader stops participating in the protocol and remains silent (this could be either seen as an attempt to impede the system to progress or the effect of a common crash failure).
Subsequently, replicas perform a leader change and abort (at $1846$s), switching back to the conservative mode.
This blocking time is similar to what a client experiences when discovering an equivocation.
At $1889$s, after finishing another $400$ consensus instances, replicas return to the fast mode,
yielding only a modest performance improvement because the weights are not yet optimized.
At $2022$s, the system self-optimizes to its best weight distribution and leader placement (using the mechanisms leveraged from \aware), reaching again latencies almost as low as experienced before.\footnote{
After the leader fails, there are fewer replicas and thus less flexibility in quorum formation. 
Since the failed leader was part of the best clique of well-connected replicas, the following configuration (after re-optimization) is slightly slower.}

\emph{Malicious leader.}
At the time $2640$s, we artificially let the current leader \texttt{eu-west-2} conduct a \emph{pre-prepare delay attack}~\cite{amir2010prime}, in which it purposely delays sending its proposal to degrade the system's performance. 
After approximately $185$s, which is the time required for a measurement round and self-optimization~\cite{aware}, \ourProtocol{} detects it is running in a sub-optimal configuration and changes replicas' weights, moving the leader to \texttt{us-east-1} to accelerate performance.

\begin{figure}[t]
	\centering
	 \begin{tikzpicture}
    \begin{axis}[
width= 5.5cm,
height=3.6cm,
font= \small,
    xlabel={Clients},
    ylabel={Throughput [kOps/s]},
    xmin=0, xmax=3200,
    ymin=0, ymax=8,
    xtick={0, 1000, 2000, 3000},
    ytick={0, 2, 4, 6,  8},
    xticklabel style={/pgf/number format/1000 sep={}},
    legend pos=south east,
    legend columns = 1,
    legend style={at={(1.85, 0)}},
    legend cell align={left},
    ymajorgrids=true,
    xmajorgrids=true,
    grid style=dashed,
]

\addplot[
    color=lb3,
    mark=square*,
    ]
    table [x=clientsTotal,y=aware3] {data/throughput-bench0-0.txt};

\addplot[
    color=lb4,
    mark=diamond*,
    ]
    table [x=clientsTotal,y=ftl] {data/throughput-bench0-0.txt};

    \addplot[
    color=lb2,
    mark=triangle*,
    ]
    table [x=clientsTotal,y=aware6] {data/throughput-bench0-0.txt};

        \addplot[
    color=lb1,
    mark=*,
    ]
    table [x=clientsTotal,y=bftsmart] {data/throughput-bench0-0.txt};

    
   \legend{AWARE $t=3$, \ourProtocol,  AWARE $t=6$,  \bftsmart/PBFT}
\end{axis}
\end{tikzpicture} 
	\caption{Throughput comparison for $n=21$ replicas.}
	\label{fig:throughput}
\end{figure}

\subsubsection{Throughput}
Although \ourProtocol{} aims to optimize latency, we conducted a simple 0/0-microbenchmark in our emulated AWS network with an increasing number of clients evenly distributed among all AWS regions while measuring the throughput of \bftsmart{}, \aware{} ($t=6$), \aware{} ($t=3$), and \ourProtocol{}.
In this experiment, we used 0-byte requests and replies to avoid the saturation of links bandwidth.
Figure~\ref{fig:throughput} presents the results, which show two main insights.
First, faster consensus instances can achieve higher throughput, provided the available network bandwidth is not exhausted.
Second, \ourProtocol{} displays a modest performance decrease compared to \aware{} ($t=3$), which uses the same quorums but offers \emph{half of our resilience}.
This difference stems from the costs of additional signatures needed to integrate lightweight forensics support into \ourProtocol{}.

\begin{figure}[t]
    \centering
    \includegraphics[width=\columnwidth]{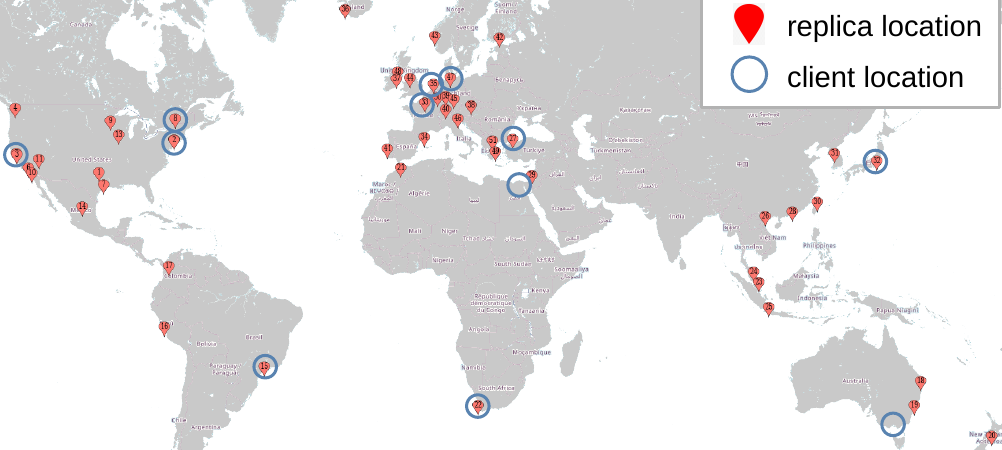}
    \caption{Map showing the locations of the 51 replicas used in our larger deployment.}
    \label{fig:wonderproxy}
\end{figure}
\begin{figure*}[t]
\centering
\begin{tikzpicture} 

    \begin{axis}[ 
     ylabel={Latency [ms]}, 
     xticklabels from table={data/eval-sim1.txt}{region},    
     ybar=0.5pt,  
         font= \small,
     bar width=4pt,
    width=1\linewidth, 
    height=4.2cm,
       xtick=data, 
       ytick = {0,100,200,300,400,500,600, 700, 800},
        ymin=0,
        ymax=900,
    ymajorgrids=true,
    yminorgrids=true,
    minor grid style={dashed,gray!10},
    minor tick num=1,
    legend style={at={(1, 1.1)},
    legend columns = 5,
    legend cell align=left
    }
    ] 
      \addplot 
       [draw = db1, 
        fill = lb1]   
        table[ 
          x=regionNr, 
          y=bftsmart    
          ] 
      {data/eval-sim1.txt}; 
      \addlegendentry{\bftsmart{}/PBFT \ \  \ourProtocol :}; 
   
            \addplot 
       [draw = dark-green, 
        fill = light-green,
        ]   
        table[ 
          x=regionNr, 
          y=final    
          ] 
      {data/eval-sim1.txt}; 
       \addlegendentry{final}; 
       \addplot 
       [draw = dark-yellow, 
        fill = light-yellow,
        ]   
        table[ 
          x=regionNr, 
          y=strong    
          ] 
      {data/eval-sim1.txt}; 
      \addlegendentry{strong};
      
      \addplot 
       [draw = dark-orange, 
        fill = light-orange,
        ]   
        table[ 
          x=regionNr, 
          y=weak    
          ] 
      {data/eval-sim1.txt}; 
      \addlegendentry{weak};
      
      \addplot 
       [draw = dark-red, 
        fill = light-red,
        ]   
        table[ 
          x=regionNr, 
          y=none   
          ] 
      {data/eval-sim1.txt}; 
      \addlegendentry{first};


\draw[black, very thick] (axis cs:0.4, 840) -- (axis cs: 0.8,840);
\node[right, black] at (axis cs:0.8 ,840) {{\small \bftsmart/PBFT at 1c}};
    \foreach \region \latency in {
    1/156.65933,
    2/182.523985,
    3/167.0301675,
    4/183.53401,
    5/186.0573375,
    6/180.9729025,
    7/188.1289575,
    8/182.338365,
    9/184.214635
    } 
    {
     \edef\temp{\noexpand\draw[black, thick] (axis cs:\region - 0.23, \latency) -- (axis cs: \region + 0.23, \latency);}
     \temp
}

\draw[gray, very thick] (axis cs:2.8, 840) -- (axis cs: 3.2,840);
\node[right, gray] at (axis cs:3.2 ,840) {{at 0.67c}};

    \foreach \region \latency in {
    1/233.819895522388,
    2/272.423858208955,
    3/249.298757462687,
    4/273.931358208955,
    5/277.697518656716,
    6/270.108809701493,
    7/280.78948880597,
    8/272.146813432836,
    9/274.94721641791
    } 
    {
     \edef\temp{\noexpand\draw[gray, thick] (axis cs:\region - 0.23, \latency) -- (axis cs: \region + 0.23, \latency);}
     \temp
}

    \end{axis}

    
\end{tikzpicture} 

\caption{Latencies of \bftsmart{} and \ourProtocol{} for $n=51$ replicas, observed from different client locations.}
\label{fig:simulator-correctables-latencies}
\end{figure*}

\begin{figure*}[t]
\centering
\begin{tikzpicture} 

    \begin{axis}[ 
     ylabel={Latency [ms]}, 
     xticklabels from table={data/eval-ftl-hs.txt}{region},    
     ybar=0.5pt,  
         font= \small,
     bar width=4pt,
    width=1\linewidth, 
    height=4.2cm,
       xtick=data, 
       ytick = {0,200, 400, 600, 800, 1000, 1200},
       yticklabel style={/pgf/number format/1000 sep={}},
        ymin=0,
        ymax=1450,
    ymajorgrids=true,
    yminorgrids=true,
    minor grid style={dashed,gray!10},
    minor tick num=1,
    legend style={at={(1, 1.1)},
    legend columns = 5,
    legend cell align=left
    }
    ] 
      \addplot 
       [draw = db2, 
        fill = lb2]   
        table[ 
          x=regionNr, 
          y=hotstuff   
          ] 
      {data/eval-ftl-hs.txt}; 
      \addlegendentry{HotStuff \ \  \ourProtocolShort HotStuff :}; 
   
            \addplot 
       [draw = dark-green, 
        fill = light-green,
        ]   
        table[ 
          x=regionNr, 
          y=final    
          ] 
      {data/eval-ftl-hs.txt}; 
       \addlegendentry{final}; 
       \addplot 
       [draw = dark-yellow, 
        fill = light-yellow,
        ]   
        table[ 
          x=regionNr, 
          y=strong    
          ] 
      {data/eval-ftl-hs.txt}; 
      \addlegendentry{strong};
      
      \addplot 
       [draw = dark-orange, 
        fill = light-orange,
        ]   
        table[ 
          x=regionNr, 
          y=weak    
          ] 
      {data/eval-ftl-hs.txt}; 
      \addlegendentry{weak};
      
      \addplot 
       [draw = dark-red, 
        fill = light-red,
        ]   
        table[ 
          x=regionNr, 
          y=none   
          ] 
      {data/eval-ftl-hs.txt}; 
      \addlegendentry{first};

\draw[black, very thick] (axis cs:1, 1340) -- (axis cs: 1.5,1340);
\node[right, black] at (axis cs:1.5 ,1340) {{\small HotStuff at 1c}};

\draw[gray, very thick] (axis cs:3, 1340) -- (axis cs: 3.5,1340);
\node[right, gray] at (axis cs:3.5 ,1340) {{\small HotStuff at 0.67c}};
    \foreach \region \latency in {
1/327.92,
2/354.42,
3/354.8,
4/337.7,
5/336.8,
6/316.38,
7/333.62,
8/336.42,
9/337.19
    } 
    {
     \edef\temp{\noexpand\draw[black, thick] (axis cs:\region - 0.23, \latency) -- (axis cs: \region + 0.23, \latency);}
     \temp
}

    \foreach \region \latency in {
1/489.43,
2/528.98,
3/529.55,
4/504.02,
5/502.68,
6/472.20,
7/497.94,
8/502.11,
9/503.26
    } 
    {
     \edef\temp{\noexpand\draw[gray, thick] (axis cs:\region - 0.23, \latency) -- (axis cs: \region + 0.23, \latency);}
     \temp
}

    \end{axis}

    
\end{tikzpicture} 

\caption{Latencies of HotStuff using \ourProtocol{} techniques for $n=51$ replicas.}
\label{fig:simulator-hs-correctables-latencies}
\end{figure*}

\subsection{Larger Deployments}

In this experiment, we assess if the performance gains of \ourProtocol{} are sustained in a different scenario with a larger number of $n=51$ replicas, approaching an expected permissioned blockchain deployment.
Since this number exceeds the available AWS regions, we sample locations from a publicly available dataset provided  by \emph{Wonderproxy}.\footnote{\url{https://wonderproxy.com/blog/a-day-in-the-life-of-the-internet/}.}
We distributed a replica in each of the chosen $51$ locations and deployed $12$ clients, as depicted in  Figure~\ref{fig:wonderproxy}. 

\subsubsection*{Setup}
For simulating this network, we use \emph{Phantom}~\cite{jansen2022co}, which employs a hybrid simulation-emulation architecture, where real, unmodified applications execute as native Linux processes within this network simulator.
Previous research has shown that Phantom can be used to faithfully evaluate the performance of BFT protocol implementations~\cite{berger2022does}.
For validity, we repeated the $n=21$ AWS experiment depicted in the previous section in this simulator and observed a marginal deviation with the results obtained from the real AWS network and Kollaps emulator (see Appendix).

In our experiment, we measure the latency speedup that \ourProtocol{} achieves in direct comparison with \bftsmart{}.
We measure both consensus latency and client end-to-end latencies using the same method described before.
Phantom bootstraps replicas and clients in their host locations with the initial protocol leader in Cape Town. As before, clients run simultaneously and send requests with a $400$-bytes payload with a randomized waiting interval of up to $4$s between two requests.
When running \ourProtocol{}, replicas are started in a configuration with optimal resilience threshold ($t=16$), but \ourProtocol{} optimizes this threshold to $t_\mathit{fast}=8$ before clients collect their measurement samples.

\subsubsection*{Results}  
Figure~\ref{fig:simulator-correctables-latencies} shows similar latency improvements as in the previous experiment for \ourProtocol{} when compared with \bftsmart{}.
The consensus latency (not shown) decreases from $350$ms in \bftsmart{} to only $88$ms in \ourProtocol, corresponding to a consensus execution speedup of $3.98\times$.
For request latencies with \emph{Final} consistency, the highest speedup observed was in Frankfurt ($1.95\times$ from $615$ms down to $314$ms), and the lowest speedup was observed in Cape Town ($1.41\times$ from $618$ms to $440$ms).

Further, the speedup increases when using the incremental consistency levels of the correctable.
For instance, in Paris, the \emph{first} level achieves a speedup of $5.52\times$, while the \emph{strong} level still provides a speedup of $4.03\times$.
The average speedup across all client locations from \bftsmart{} to \ourProtocol{}' \emph{final} level is $1.83\times$ and becomes incrementally higher for the speculative levels \emph{strong} ($2.70\times$), \emph{weak} ($2.99\times$) and \emph{first} ($3.23\times$).
For comparison, the speedup that \bftsmart{} would achieve if the speed of network links approximated the speed of light is roughly $2.5\times$.
These results show that using smaller weighted quorums is effective for reducing latency, with values similar to the expected latency of an optimal protocol (\bftsmart/PBFT) using speed-of-light network links.

\subsection{\ourProtocol{}-flavored HotStuff}

The principle of improving quorums introduced in \ourProtocol{} is general and can be used in other BFT SMR protocols to decrease latency.
For example, our techniques can be directly applied to speedup agreement in multi-leader protocols~\cite{scalabilitymadesimple}, as long as the leaders are selected only among the best-connected replicas, or even in protocols providing additional guarantees such as fair ordering~\cite{zhang20pompe}.
Here, we experimentally demonstrate this aspect by applying the \ourProtocol{} transformation to HotStuff~\cite{hotstuff19}.

Compared to \bftsmart{} and PBFT, HotStuff uses an agreement pattern with one additional phase and achieves a linear communication complexity by letting the leader collect and distribute quorum certificates in each phase.
It results in $7$ communication steps per consensus instance instead of $3$ as required by \bftsmart{}/PBFT.
This design makes the overall system's latency even more sensitive to how fast agreement can be achieved, which depends on the speed at which a HotStuff leader can succeed in collecting quorum certificates. 
\ourProtocolShort HotStuff selects a configuration where the leader communicates fast with a set of well-connected replicas granted high voting power.

\subsubsection*{Setup}
We use a prediction model of the HotStuff protocol\footnote{This model is similar to the one \aware{} uses to anticipate the effect of weight and leader changes during its self-optimization~\cite{aware}.} to simulate \ourProtocolShort HotStuff and the original HotStuff protocol running on the same latency map of $n=51$ replicas used before (Figure~\ref{fig:wonderproxy}).
These simulations compute the achievable latencies for different consistency levels of \ourProtocolShort HotStuff, the latency of the HotStuff, and the hypothetical latencies of HotStuff with speed-of-light links.

\subsubsection*{Results}
Our results show that \ourProtocolShort HotStuff significantly optimizes HotStuff's consensus latency (see Figure~\ref{fig:simulator-hs-correctables-latencies}) from $853$ms to only $177$ms in \ourProtocolShort HotStuff.
It corresponds to a consensus execution speedup of $4.82\times$, achieved by the incorporation of weights, optimal leader placement, and the use of smaller quorums.
For client request latencies with \emph{final} consistency, the highest speedup observed was in Frankfurt ($2.84\times$) and lowest in Cape Town ($2.07\times$). 

Like before, 
the speedup increases with lower consistency levels.
For instance, in Paris, the \emph{first} level achieves a speedup of $6.03\times$ (from 1141~ms down to 189~ms), while the \emph{strong} level still provides a speedup of $4.71\times$ ($242$~ms).
The average speedup across all client locations from HotStuff to  \ourProtocolShort{} HotStuff's \emph{final} level is $2.56\times$ and becomes incrementally higher for the speculative levels \emph{strong} ($3.69\times$), \emph{weak} ($4.05\times$) and \emph{first} ($4.74\times$).


\section{Related Work}

\paragraph{Adaptivity in BFT SMR}
Making BFT protocols adaptive to their environment has been studied in multiple works~\cite{aublin2013rbft, bahsoun2015making, liu2017leader, eischer2018latency, carvalho2018dynamic, silva2021threat, chiba2022network, nischwitz2022raising,boltdumbo}.
RBFT monitors system performance under redundant leaders to prevent a faulty leader from degrading performance~\cite{aublin2013rbft}.
Other approaches propose optimizing the leader selection (e.g., \cite{liu2017leader, eischer2018latency}),
adaptively switching consensus algorithms~\cite{bahsoun2015making, carvalho2018dynamic,boltdumbo}, strengthening the protocol by reacting to perceived threat level changes~\cite{silva2021threat}, being network-agnostic (tolerating a higher threshold in synchronous networks)~\cite{blum2020network}, or adapting the state transfer strategy to the available network bandwidth~\cite{chiba2022network}.
Bolt-Dumbo~\cite{boltdumbo} runs a fast quorum-based protocol in synchronous periods and falls back to an asynchronous consensus otherwise.
In contrast, \ourProtocol{} is applicable to most BFT protocols and accelerates planetary-scale Byzantine consensus by making both system configuration (leader and replica weights) \emph{and} threshold adaptive without impacting resilience through the integration of BFT protocol forensics.

\paragraph{Geographically-distributed SMR}
Various works studied the improvement of SMR for WANs~\cite{mao2008mencius, mao2009towards, wester2009tolerating, veronese2010ebawa, amir10steward, coelho18geopaxos, numakura2019evaluation, eischer2020low, eischer2020resilient, neiheiser2020fireplug, gupta2020resilientdb, yahyaoui2024tolerating}.
Mencius, one of the earliest of these works, optimizes WAN performance using a rotating leader scheme that allows clients to pick their geographically closest replica as its leader~\cite{mao2008mencius}, but
tolerating only crash faults. EBAWA uses the same rotating leader technique together with trusted components on each replica to tolerate Byzantine replicas in a protocol with the same number of communication steps as Mencius~\cite{veronese2010ebawa}.
Steward proposes a hierarchical, two-layered replication architecture.
Regional groups within a system site run Byzantine agreements, and these replication groups are then connected with a CFT protocol~\cite{amir10steward}.
Fireplug~\cite{neiheiser2020fireplug} later adapts this hierarchical architecture for efficient geo-replication of graph databases by composing multiple BFT-SMaRt groups.
GeoBFT~\cite{gupta2020resilientdb} assumes regional clusters and employs hierarchical consensus to first replicate a client transaction in its local replication group, and afterwards the transactions of all local groups are shared globally, then ordered and executed. A similar approach was used in~\cite{yahyaoui2024tolerating}, which employs the Damysus protocol~\cite{decouchant2022damysus} to decide on superblocks in the upper consensus layer. In hierarchical approaches, it is assumed that the number of failures is bounded for each of the clusters.

WHEAT optimizes BFT SMR latency by incorporating weighted replication and tentative executions~\cite{wheat}, while \aware{} enriches WHEAT through self-monitoring capabilities and dynamic optimization by adjusting weights and leader position~\cite{aware}.
By integrating lightweight forensics, \ourProtocol{} can safely use smaller consensus quorums and accelerate Byzantine consensus even further than \aware, thus mastering the resilience-performance trade-off that limits \aware's performance.

\paragraph{Fast or Speculative BFT}
It has been shown that having additional redundancy (and using a less-than-optimal resilience threshold) can be efficiently utilized to develop ``fast'' consensus variants, i.e., two-step Byzantine consensus~\cite{martin2006fast, howard2021fast, kuznetsov2021revisiting}, or even one-step asynchronous Byzantine consensus for scenarios that are  contention-free~\cite{friedman05simple,song2008bosco}.
DuoBFT~\cite{arun2020duobft} uses the hybrid and Byzantine fault models where clients can choose the favored model for each command.
Since hybrid commits take fewer communication steps and use smaller quorums than BFT commits, clients benefit from low-latency commits in the hybrid model.
In comparison, \ourProtocol{} extends a latency- and resilience-optimal protocol (PBFT) to significantly improve latency without requiring more than $3t+1$ replicas.

Some BFT protocols propose mechanisms based on speculation to accelerate the overall protocol when running in a ``common case'' scenario (often assuming the absence of failures or congestion)~\cite{pbft, kotla2007zyzzyva, wester2009tolerating, Gupta2019Proof}.
A form of server-side speculation technique was initially proposed by PBFT as \textit{tentative execution}, in which replicas execute and respond to requests directly after the \textsc{prepare} stage~\cite{pbft}.
In the Proof-of-Execution protocol, server-side speculation after the \textsc{prepare} stage was revisited, formally specified, and proved correct~\cite{Gupta2019Proof}. 
\textit{Tentative executions} are orthogonal to the techniques explored in this paper and are prototypically implemented and explored by the WHEAT protocol~\cite{wheat} which AWARE and \ourProtocol{} are extending.

Zyzzyva proposes a form of \emph{speculative execution} in which replicas execute requests directly after receiving a proposal from the leader~\cite{kotla2007zyzzyva}. 
Yet Zyzzyva requires a predictable and stable network that is uncommon in geo-distributed deployments and shows quickly degrading performance in case of failures, as it necessitates collecting responses from \emph{all} replicas within some time window to complete a request in a single phase~\cite{singh2008bft}.
PBFT-CS refines PBFT with client-side speculation.
Clients send subsequent requests after predicting a response to an earlier request without waiting for the earlier request to commit---however, clients need to track and propagate the dependencies between requests~\cite{wester2009tolerating}.

In contrast, \ourProtocol{} incorporates a novel technique for client-side speculation to allow applications to work on correctable results (as first proposed in~\cite{guerraoui2016incremental}) obtained from the replicated state machine by proposing \textit{Byzantine correctables} which offer increasing consistency guarantees to clients.

\paragraph{Accountability in BFT}
BFT forensics is a technique for analyzing safety violations in BFT protocols after they happened~\cite{bftforensics}, yielding results such as that at least $t+1$ culprits can be identified in case of an equivocation (with the accountability of up to $\frac{2n}{3}$ replicas that may be Byzantine).
Polygraph is an accountable Byzantine consensus algorithm tailored for blockchain applications that allow the punishment of culprits (e.g., via stake slashing) in case of equivocations~\cite{civit2021polygraph}.
A simple transformation to obtain an accountable Byzantine consensus protocol from any Byzantine consensus protocol has been proposed in~\cite{civit2022easy}.
The Basilic class of protocols solves consensus with $n \leq 3t + d + 2q$ replicas tolerating~$t$~general Byzantine failures, $d$ deceitful failures (that violate safety---the ones that forensics can identify), and~$q$~benign failures~\cite{ranchal2022basilic}.
Basilic's resilience has been proven optimal. 
FireLedger~\cite{buchnik2020} proposes a high-throughput blockchain consensus protocol in which the last $t + 1$ blocks of every replica’s blockchain are considered \textit{tentative} and replicas verify the correctness (finality) of these blocks later on. A malicious proposer can be detected using a proof and is removed from the system through a recovery  procedure.
IA-CCF~\cite{shamis22iaccf} shows that logging all messages exchanged in PBFT-based blockchain makes it possible to identify any misbehaving replicas in case of equivocations.

\ourProtocol{} does not differentiate failures or require a blockchain, employing instead a light version of the BFT forensics protocol of~\cite{bftforensics} to identify a limited number of equivocators in fast mode.

\section{Conclusion}
\label{sect:conclusion}

\ourProtocol{} accelerates planetary-scale Byzantine consensus by combining weighted replication with lightweight BFT forensics to safely underestimate the resilience threshold, using faster quorums to drive consensus decisions.
We showed how to obtain \ourProtocol{} from \aware{} by 
utilizing BFT forensics techniques in a novel way: as a protective countermeasure against attacks.
Notably, \ourProtocol{} always achieves linearizability and liveness under the optimal resilience threshold, even when quorums are formed using the fast threshold. 
Our evaluation results indicate that latency benefits are substantial, i.e., achieving a speedup of  $1.87\times$ over \bftsmart/PBFT. 
Our methodology is a transformation applicable in other BFT protocols, improving their speed under geographical dispersion. 
We showed that if a protocol's agreement pattern consists of more communication steps (like in HotStuff), it results in even greater benefits ($2.56\times$ speedup). 

We also employed client-side speculation, allowing the application to choose a relaxed consistency level from the client-replica contract on the granularity level of individual operations. 
This type of optimization can be a good fit for time-sensitive/low-risk transactions (e.g., micro-payments) as they can benefit from up to $6\times$ speedups. 
Moreover, we think there are interesting use cases for what applications can do with requests before they have been finalized.
For instance, sequential consistency (called ``weak'' in our evaluations) might suffice in applications where operations work on non-shared objects (e.g., transferring a coin from an account $A$ to $B$ in a UTXO-based account model~\cite{nakamoto2008bitcoin}), in which operations issued by different clients rarely conflict.
Furthermore, even the most speculative level ``first'' could be used to speed up an application like a BFT/blockchain-based name resolver.
Such a service could resolve a name, and then the client uses the intermediate (correctable result) to pre-fetch content from a web server.
If the obtained result is incorrect, then the client would need to throw away pre-loaded content and initiate a new query to the correct server (the programming interfaces in the correctable define such behavior).
Our evaluation results demonstrate that the consistency level ``strong'' (which preserves linearizability under the assumption of $f \leq t_{fast}$) already achieves high speedups and might present an interesting sweet spot for low-risk transaction settlement.

\section*{Acknowledgments}
We thank the anonymous Middleware reviews and our shepherd, Mohammad Sadoghi, for their insightful comments.
This work was supported by FCT through the ThreatAdapt (FCT-FNR/0002/2018) and \href{https://doi.org/10.54499/2022.08431.PTDC}{SMaRtChain (2022. 08431.PTDC)} projects, and the \href{https://doi.org/10.54499/UIDB/00408/2020}{LASIGE Research Unit (UIDB/00408/2020} and \href{https://doi.org/10.54499/UIDP/00408/2020}{UIDP/00408/2020)}.
It has also been funded by Deutsche Forschungsgemeinschaft (DFG, German Research Foundation) grant number 446811880 (BFT2Chain).





\bibliographystyle{ACM-Reference-Format}
\bibliography{main}

%
%
%

\appendix

\section*{Appendix}

In this appendix, we present the full correctness proof of the \ourProtocol{} and the experimental results about the validity of our simulation and emulation scenarios when compared with a real network.
In particular, in this appendix, we prove the following theorems.

\vspace{3mm}
\noindent
\textbf{Theorem 1.}
\emph{If an operation $o$ is finalized in \mbox{$i$-th} position of the decision log, then no client observes an operation $o'\neq o$ in this position of the decision log.}

\vspace{3mm}
\noindent
\textbf{Theorem 2.}
\emph{An operation issued by a correct client will eventually be finalized.}


\section*{Correctness of \ourProtocol}
\label{sec:correctness}

In the following, we prove the correctness of \ourProtocol{} transformation, as summarized in Figure~\ref{fig:ftl-description}. 
In particular, we prove it preserves the safety and liveness of its underlying protocol, \aware~\cite{aware}, with up to $t$ failures.

To abstract the exact service being implemented on top of \ourProtocol{}, our proofs consider the replicated decision log abstraction as described in \S 2.1, in which every operation needs to be allocated consistently in a given position/slot of the log.
We say an operation is \emph{finalized} when the client that issued it knows it was durably executed and will never be reverted in the system.
The exact number of replies required for finalizing an operation is defined in Rule~\clientRule{\footnotesize C2}, in Figure~\ref{fig:ftl-description}.

\subsection*{Safety}

Instead of only proving all operations are executed in total order, we have to prove all clients observe operations in total order, which ensures linearizability~\cite{herlihy1990linearizability}.
To prove that, we start by showing that an operation finalized in some position of the decision log in a mode of operation is kept in that position after a mode switch.

\begin{proposition}\label{consfast}
Let $o$ be an operation finalized in \emph{conservative} mode in the \mbox{$i$-th} position of the decision log.
After the system switches to subsequent \emph{fast} mode, $o$ will still be the \mbox{$i$-th} operation executed in the system.
\end{proposition}

\begin{proof}
Assume, without loss of generality, that the system switches from \emph{conservative} to \emph{fast} mode right after executing its \mbox{$j$-th} operation, with $j \geq i$.
In this case, all operations ordered after the switch will be finalized in positions after~$j$ (Rule~\serverRule{\footnotesize S3} of Figure~\ref{fig:ftl-description}), and thus the effect of previously ordered operations (including $o$) will be maintained.
\end{proof}

\begin{lemma}\label{fastcons}
Let $o$ be an operation finalized in \emph{fast} mode in the \mbox{$i$-th} position of the decision log.
After a synchronization phase that switches the system to the \emph{conservative} mode, $o$ will still be the \mbox{$i$-th} operation executed in the system.
\end{lemma}

\begin{proof}
Let's assume, for the sake of contradiction, that an operation $o$ finalized in the \mbox{$i$-th} position of the decision log in fast mode does not appear in this position after a synchronization phase.
This can happen for one of two reasons:

\begin{enumerate}
\item \emph{Operation $o$ was erased, making the \mbox{$i$-th} position empty.}
This can only happen if there is no correct replica in the intersection between the quorum of $n-t_\mathit{fast}-1$ replicas that informed the client about the finalization of $o$ and the quorum of $n-t$ replicas that informed the new leader about the requests ordered in fast mode during the synchronization phase.
This is not possible since these quorums intersect in $(n-t_\mathit{fast}-1) + (n-t) - n = \frac{3t}{2}$ replicas\footnote{This value is obtained by using $t_\mathit{fast} = \frac{t}{2}$ and $n = 3t+1$.}, which is clearly bigger than $t$, for $t \geq 2$ (when the fast mode brings benefits).

\item \emph{Operation $o$ was replaced by another operation $o'$ in the \mbox{$i$-th} position.}
This happens only if the number of replicas reporting $o'$ as prepared is higher than the number of replicas reporting $o$ in the synchronization phase quorum of $n-t$ replicas (Rule \serverRule{\footnotesize S6} step 2 of the transformation in Figure~\ref{fig:ftl-description}).
Since $o$ was finalized, at least $C = n-t_\mathit{fast}-1-t$ \emph{correct replicas} informed the client about the execution of $o$.
Under an equivocation scenario, operation $o'$ would be reported by at most $C' = n - C + t = t+t_\mathit{fast}+1$ replicas, including the equivocators that participated in the decision of $o$~and~$o'$.
Notice that the $t_\mathit{fast}+1$ replicas that maliciously voted in the preparation of the two requests can be detected by the lightweight forensics protocol used in the synchronization phase (Rule \serverRule{\footnotesize S6} step 1 of the transformation).
Therefore, these replicas will be ignored in the synchronization phase quorum (Rule~\serverRule{\footnotesize S6} step 2) (and later expelled from the system---Rule~\serverRule{\footnotesize S6} steps 3 and 4), ensuring no more than $t$ replicas report $o'$ in this quorum.
Consequently, even in this worst-case scenario, the new leader will always see $n-2t > t$ replicas reporting $o$, being thus the most common value appearing in the reported values.
This result will make $o$ be kept in the \mbox{$i$-th} position of the log after a synchronization phase.
\end{enumerate}
\end{proof}

The following theorem proves \ourProtocol{}' main safety property: finalized operations are observed in the same position of the decision log by every correct client.

\begin{theorem}
If an operation $o$ is finalized in \mbox{$i$-th} position of the decision log, then no client observes an operation $o'\neq o$ in this position of the decision log.
\end{theorem}

\begin{proof}
We start by observing that, according to Rule~\clientRule{\footnotesize C2} of the transformation (Figure~\ref{fig:ftl-description}), a client accepts an operation result as finalized only if the number of matching $\textsc{reply}$ messages \emph{in the same mode} satisfies certain quorum sizes.
Let $mode(o)$ be the mode of operation in which $o$ was finalized, and assume, for the sake of contradiction, that an operation $o' \neq o$ appears as finalized in the \mbox{$i$-th} position for some correct client. 
\begin{figure*}[t!]
\centering
\begin{tikzpicture}
    \begin{axis}[ 
     ylabel={Latency [ms]}, 
     xticklabels from table={data/eval-real-aws.txt}{region},    
     ybar=0.5pt,  
         font= \small,
     bar width=4pt,
    width=0.8\linewidth, 
    height=4.2cm,
       xtick=data, 
            xticklabel style={rotate=10},
       ytick = {0,100,200,300,400,500,600, 700, 800},
        ymin=0,
        ymax=1000,
    ymajorgrids=true,
    yminorgrids=true,
    minor grid style={dashed,gray!10},
    minor tick num=1,
    legend style={at={(1, 1.25)},
    legend columns = 6,
    legend cell align=left
    }
    ] 
      \addplot 
       [draw = db1, 
        fill = lb1]   
        table[ 
          x=regionNr, 
          y=bftsmart    
          ] 
      {data/eval-real-aws.txt}; 
      \addlegendentry{\bftsmart{}}; 

          \addplot 
       [draw = db3, 
        fill = lb3]   
        table[ 
          x=regionNr, 
          y=aware    
          ] 
      {data/eval-real-aws.txt}; 
      \addlegendentry{AWARE \ \ \ \ \ourProtocol:}; 
   
            \addplot 
       [draw = dark-green, 
        fill = light-green,
        ]   
        table[ 
          x=regionNr, 
          y=final    
          ] 
      {data/eval-real-aws.txt}; 
       \addlegendentry{final}; 
       \addplot 
       [draw = dark-yellow, 
        fill = light-yellow,
        ]   
        table[ 
          x=regionNr, 
          y=strong    
          ] 
      {data/eval-real-aws.txt}; 
      \addlegendentry{strong};
      
      \addplot 
       [draw = dark-orange, 
        fill = light-orange,
        ]   
        table[ 
          x=regionNr, 
          y=weak    
          ] 
      {data/eval-real-aws.txt}; 
      \addlegendentry{weak};
      
      \addplot 
       [draw = dark-red, 
        fill = light-red,
        ]   
        table[ 
          x=regionNr, 
          y=none   
          ] 
      {data/eval-real-aws.txt}; 
      \addlegendentry{first};

  \filldraw[red, semitransparent] (axis cs:2.5, 925) circle (2pt);
\node[right, red] at (axis cs:2.55 ,920) {{Simulation (Phantom)}};
    \foreach \region \latency in {
1/718.375641,
2/681.206599,
3/790.861935,
4/601.251740,
5/636.898955,
6/796.018700,
7/649.639637
    } 
    {
     \edef\temp{\noexpand
     \filldraw[red, semitransparent] (axis cs:\region - 0.225, \latency) circle (2pt);
     }
     \temp
}

\foreach \region \latency in {
1/550.853548,
2/494.906795,
3/603.883581,
4/587.434532,
5/461.910367,
6/604.363138,
7/495.250022
    } 
    {
     \edef\temp{\noexpand
       \filldraw[red, semitransparent] (axis cs:\region - 0.135, \latency) circle (2pt);
     }
     \temp
}

\foreach \region \latency in {
1/375.2574126,
2/358.1826183,
3/464.0029354,
4/484.9974852,
5/346.8338640,
6/424.0227085,
7/360.1603646
    } 
    {
     \edef\temp{\noexpand
       \filldraw[red, semitransparent] (axis cs:\region - 0.045, \latency) circle (2pt);
     }
     \temp
}

\foreach \region \latency in {
1/365.59828401,
2/259.57528016,
3/347.32533875,
4/373.22861296,
5/233.18132537,
6/411.78378069,
7/304.96942966
    } 
    {
     \edef\temp{\noexpand
       \filldraw[red, semitransparent] (axis cs:\region + 0.045, \latency) circle (2pt);
     }
     \temp
}

\foreach \region \latency in {
1/341.45830760,
2/245.99660582,
3/333.48725177,
4/314.95237570,
5/176.72707716,
6/403.64592467,
7/244.60318863
    } 
    {
     \edef\temp{\noexpand
       \filldraw[red, semitransparent] (axis cs:\region + 0.135, \latency) circle (2pt);
     }
     \temp
}

\foreach \region \latency in {
1/323.077208,
2/234.439159,
3/322.969549,
4/302.526644,
5/168.522284,
6/382.948299,
7/237.735036
    } 
    {
     \edef\temp{\noexpand
       \filldraw[red, semitransparent] (axis cs:\region + 0.225, \latency) circle (2pt);
     }
     \temp
}

\draw[blue, very thick] (axis cs:0.5, 925) -- (axis cs: 0.6,925);
\node[right, blue] at (axis cs:0.6 ,920) {{Emulation (Kollaps)}};

    \foreach \region \latency in {
1/759.171549,
2/704.137622,
3/820.850133,
4/639.343489,
5/667.923811,
6/829.399950,
7/698.394731
    } 
    {
     \edef\temp{\noexpand
     \draw[blue, thick] (axis cs:\region - 0.26, \latency) -- (axis cs: \region - 0.19, \latency);
     }
     \temp
}
    \foreach \region \latency in {
1/551.44731520,
2/493.51039384,
3/607.34424417,
4/595.31833867,
5/469.30374194,
6/606.87316891,
7/496.37738920
    } 
    {
     \edef\temp{\noexpand
     \draw[blue, thick] (axis cs:\region - 0.17, \latency) -- (axis cs: \region - 0.10, \latency);
     }
     \temp
}

    \foreach \region \latency in {
1/390.657195259,
2/361.132595461,
3/457.975911662,
4/468.962742138,
5/333.809703151,
6/433.798679602,
7/364.250823775
    } 
    {
     \edef\temp{\noexpand
     \draw[blue, thick] (axis cs:\region - 0.08, \latency) -- (axis cs: \region - 0.01, \latency);
     }
     \temp
}

    \foreach \region \latency in {
1/360.6038822,
2/274.7308076,
3/356.4649645,
4/373.6380686,
5/235.8758263,
6/408.9413191,
7/309.8996737
    } 
    {
     \edef\temp{\noexpand
     \draw[blue, thick] (axis cs:\region + 0.01, \latency) -- (axis cs: \region + 0.08, \latency);
     }
     \temp
}

    \foreach \region \latency in {
1/349.452748,
2/256.301513,
3/335.604739,
4/329.841137,
5/194.594427,
6/400.781481,
7/265.313158
    } 
    {
     \edef\temp{\noexpand
     \draw[blue, thick] (axis cs:\region + 0.1, \latency) -- (axis cs: \region + 0.17, \latency);
     }
     \temp
}

    \foreach \region \latency in {
1/311.344505,
2/239.157128,
3/319.662928,
4/318.954179,
5/183.615525,
6/359.207406,
7/255.791032
    } 
    {
     \edef\temp{\noexpand
     \draw[blue, thick] (axis cs:\region + 0.19, \latency) -- (axis cs: \region + 0.26, \latency);
     }
     \temp
}

    \end{axis}

    
\end{tikzpicture} 

\caption{Comparison of clients' observed end-to-end latencies for protocol runs with \bftsmart, AWARE and \ourProtocol{} in different network environments: real, emulated, and simulated. The client results are averaged over all regions per continent.}
\label{fig:validity-of-networks}
\end{figure*}

To show correct clients cannot observe different operations $o$ and $o'$ in \mbox{$i$-th} position of the system history, we need to consider all possible combinations of modes for $o$ and $o'$.
We start by considering the cases in which both operations were finalized in the same mode without any switch in the system mode between their executions.
There are two cases to consider:

\begin{enumerate}
\item $mode(o) = mode(o') = \mathit{conservative}$: 
if both operations were executed in the conservative mode, then the total order of operations ensured by \aware's state machine replication algorithm~\cite{aware,sousa2012byzantine} makes it impossible for different clients to observe finalized operations $o$ and $o'$ in the same position in the system history.

\item $mode(o) = mode(o') = \mathit{fast}$: 
if both operations were ordered in fast mode, then $f > t_\mathit{fast}$ malicious replicas (including the leader) can lead to different correct replicas deciding different operations for the same position $i$ of the decision log.
In this case, we have to show that clients waiting for $n-t_\mathit{fast}-1$ matching replies is enough to ensure they cannot observe two finalized operations in the same position.
This holds because the size of the intersection of any two reply quorums is $(n-t_\mathit{fast}-1) + (n-t_\mathit{fast}-1) - n = n - t - 2$, which is bigger than $t$ for any $t \geq 2$ (when the fast mode can be used).
This means correct clients will not observe two finalized operations in the same decision log position in our system model.
\end{enumerate}

Now, we need to consider the cases in which there was exactly one mode switch between the finalization of $o$~and~$o'$.
There are two cases to consider:

\begin{enumerate}
\item $mode(o) = \mathit{conservative}$ and $mode(o') = \mathit{fast}$: 
Proposition~\ref{consfast} states that finalized operation $o$ position cannot be altered in a conservative-to-fast switch, i.e., it will still appear as the \mbox{$i$-th} operation executed, and thus operation $o'$ cannot appear in position $i$.

\item $mode(o) = \mathit{fast}$ and $mode(o') = \mathit{conservative}$: 
Lemma~\ref{fastcons} states that if $o$ was finalized in fast mode, a subsequent synchronization phase that switches the system to conservative mode keeps the $o$ in the \mbox{$i$-th} position of the decision log, making thus impossible for another operation $o'$ to be finalized in this position.

\end{enumerate}

Lastly, we need to consider all the cases above, but in which multiple mode switches happen between the finalization of $o$ and $o'$.
Proposition~\ref{consfast} and Lemma~\ref{fastcons} ensure a finalized operation is preserved in the system state even after a mode switch.
Consequently, by applying induction arguments, it is easy to see that the number of switches between the finalization of $o$ and $o'$ does not change the fact $o$ will always remain the \mbox{$i$-th} operation in the decision log.
\end{proof}

\subsection*{Liveness} 

As defined in \S 3, SMR liveness comprises the guarantee that operations issued by correct clients are eventually finalized.
Proving the liveness of BFT protocols is generally perceived as considerably more intricate than proving their safety~\cite{bftliveness}.
However, this distinction does not apply to \ourProtocol{} because our transformation relies upon the underlying protocol (\aware, a variant of the well-known BFT-SMaRt) for ensuring this property.
The following theorem defines the liveness of \ourProtocol.

\begin{theorem}
An operation issued by a correct client will eventually be finalized.
\end{theorem}

\begin{proof}
To argue about that, suppose a correct client $c$ sends operation $o$ to the replicas.
As stated before, an operation is finalized if the client observes it was executed in a certain position by sufficiently many replicas in the same mode (Rule~\clientRule{\footnotesize C2} of Figure~\ref{fig:ftl-description}).
Again, we have to consider the two modes of \ourProtocol{}:

\begin{enumerate}

\item \emph{Conservative mode:} 
If the leader is correct and there is sufficient synchrony, then a batch containing $o$ will be eventually decided through \textsc{aware} (Rule~\serverRule{\footnotesize S1}).
In case of a faulty leader or asynchrony, the timers associated with $o$ on correct replicas will expire, and the synchronization phase will be executed until a correct leader forces replicas to decide a batch containing $o$ (potentially after GST). 
At this point, at least $n-t$ correct replicas will send matching replies, which are collected by $c$ until it has a weighted response quorum~\cite{wheat} to finalize $o$.

\item \emph{Fast mode:} 
In fast mode, before GST no liveness can be ensured, and the system will switch back to the conservative mode. 
After GST, we have to consider two cases:

\begin{enumerate}

\item Case $f \leq t_\mathit{fast}$ and correct leader: 
This case is analogous to the conservative mode because \textsc{aware}$\star$ solves consensus with up to $t_\mathit{fast}$ faulty replicas.

\item Case $t_\mathit{fast} < f \leq t$ or faulty leader: 
In this case, \textsc{aware}$\star$ \emph{might not} be finished, and the timers associated with $o$ (Rule~\serverRule{\footnotesize S1}) will expire in correct replicas. 
This will cause replicas to initiate the synchronization phase (Rule~\serverRule{\footnotesize S4}), switching to the conservative mode, in which the request will be ordered and finalized, as explained above.
Alternatively, $f$ Byzantine replicas might participate in consensus but not reply to $c$, which will never be able to collect $n-t_\mathit{fast}-1$ matching replies. 
In this case, the client periodically reattempts to confirm the result of $o$ by checking the log of decisions (Rule~\clientRule{\footnotesize C2}).
This can be repeated until the next periodic checkpoint is formed to verify in which position on the decision log $o$ appears (Rule~\serverRule{\footnotesize S2}).
This will eventually happen because if $o$ is not ordered successfully, a timeout will trigger at the replicas, causing them to initiate the synchronization phase and switch the protocol to the conservative mode.
The liveness argument from the conservative mode then eventually holds for $o$.
\end{enumerate}
\end{enumerate}
\end{proof}

\section*{Validation of Emulated/Simulated Networks}
\label{subsec:validation-of-networks}

In this section, we compare the results of the experiment conducted in \S 6.1.1 for which we used the real AWS cloud infrastructure with supplementary experiments that use an emulated and simulated network environment that mimic the AWS infrastructure by using latency statistics from cloudping.
For these network environments, we use state-of-the-art network emulation and simulation tools, namely Kollaps~\cite{gouveia2020kollaps} and Phantom~\cite{jansen2022co}.
The repeated experiments should give some insights into how close real network characteristics can be modeled using emulation and simulation tools.
A threat to validity is that the network statistics average latency observations over a larger period (i.e., a year). 
In contrast, when conducting experimental runs, short-time fluctuations might make individual network links appear faster or slower than usual, which can impact the speed at which certain quorums are formed.
Nevertheless, we are convinced that using the network tools and latency statistics creates reasonably realistic networks that can be used to validate the latency improvements of the quorum-based protocols we are working with.

In Figure~\ref{fig:validity-of-networks}, we contrast protocol runs for \bftsmart{}, AWARE, and \ourProtocol{} on a real network, as well as in the Kollaps-based and Phantom-based networks.
\balance
On average, we observed latencies that were $1.5\%$ higher in the emulated network than on the real AWS network. Respectively, the simulated network yielded latency results that were, on average, $0.8\%$ lower than the real network.

\end{document}